\documentclass[traditabstract, longauth]{aa} 
\usepackage{graphicx}
\usepackage{fixltx2e}
\usepackage{txfonts}
\usepackage[breaklinks, colorlinks, citecolor=blue]{hyperref}
\usepackage{natbib}
\bibpunct{(}{)}{;}{a}{}{,} 
\newcommand{\thetafive}{\theta_{500}}

\newcommand{\Yscaled}{\tilde{Y}_{500}}
\newcommand{\sigscaled}{\tilde{\sigma}_{\thetafive}}

\newcommand{\YM}{Y_{\rm M}}
\newcommand{\aM}{\alpha_{\rm M}}

\newcommand{\Mhalo}{M_{\rm h}}

\def\setsymbol#1#2{\expandafter\def\csname #1\endcsname{#2}}
\def\getsymbol#1{\csname #1\endcsname}

\def\Planck{{\it Planck\/}}

\newbox\tablebox    \newdimen\tablewidth
\def\leaderfil{\leaders\hbox to 5pt{\hss.\hss}\hfil}
\def\endPlancktable{\tablewidth=\columnwidth 
    $$\hss\copy\tablebox\hss$$
    \vskip-\lastskip\vskip -2pt}

\def\tablenote#1 #2\par{\begingroup \parindent=0.8em
    \abovedisplayshortskip=0pt\belowdisplayshortskip=0pt
    \noindent
    $$\hss\vbox{\hsize\tablewidth \hangindent=\parindent \hangafter=1 \noindent
    \hbox to \parindent{$^#1$\hss}\strut#2\strut\par}\hss$$
    \endgroup}
\def\doubleline{\vskip 3pt\hrule \vskip 1.5pt \hrule \vskip 5pt}

\def\L2{\ifmmode L_2\else $L_2$\fi}

\def\DeltaT{\ifmmode \Delta T\else $\Delta T$\fi}
\def\deltat{\ifmmode \Delta t\else $\Delta t$\fi}
\def\fknee{\ifmmode f_{\rm knee}\else $f_{\rm knee}$\fi}
\def\Fmax{\ifmmode F_{\rm max}\else $F_{\rm max}$\fi}
\def\solar{\ifmmode{\rm M}_{\mathord\odot}\else${\rm M}_{\mathord\odot}$\fi}
\def\Msolar{\ifmmode{\rm M}_{\mathord\odot}\else${\rm M}_{\mathord\odot}$\fi}
\def\Lsolar{\ifmmode{\rm L}_{\mathord\odot}\else${\rm L}_{\mathord\odot}$\fi}

\def\inv{\ifmmode^{-1}\else$^{-1}$\fi}
\def\mo{\ifmmode^{-1}\else$^{-1}$\fi}
\def\sup#1{\ifmmode ^{\rm #1}\else $^{\rm #1}$\fi}
\def\expo#1{\ifmmode \times 10^{#1}\else $\times 10^{#1}$\fi}
\def\,{\thinspace}
\def\lsim{\mathrel{\raise .4ex\hbox{\rlap{$<$}\lower 1.2ex\hbox{$\sim$}}}}
\def\gsim{\mathrel{\raise .4ex\hbox{\rlap{$>$}\lower 1.2ex\hbox{$\sim$}}}}

\def\simprop{\mathrel{\raise .4ex\hbox{\rlap{$\propto$}\lower 1.2ex\hbox{$\sim$}}}}
\def\deg{\ifmmode^\circ\else$^\circ$\fi}
\def\pdeg{\ifmmode $\setbox0=\hbox{$^{\circ}$}\rlap{\hskip.11\wd0 .}$^{\circ}
          \else \setbox0=\hbox{$^{\circ}$}\rlap{\hskip.11\wd0 .}$^{\circ}$\fi}
\def\arcs{\ifmmode {^{\scriptstyle\prime\prime}}
          \else $^{\scriptstyle\prime\prime}$\fi}
\def\arcm{\ifmmode {^{\scriptstyle\prime}}
          \else $^{\scriptstyle\prime}$\fi}
\newdimen\sa  \newdimen\sb
\def\parcs{\sa=.07em \sb=.03em
     \ifmmode \hbox{\rlap{.}}^{\scriptstyle\prime\kern -\sb\prime}\hbox{\kern -\sa}
     \else \rlap{.}$^{\scriptstyle\prime\kern -\sb\prime}$\kern -\sa\fi}
\def\parcm{\sa=.08em \sb=.03em
     \ifmmode \hbox{\rlap{.}\kern\sa}^{\scriptstyle\prime}\hbox{\kern-\sb}
     \else \rlap{.}\kern\sa$^{\scriptstyle\prime}$\kern-\sb\fi}
\def\ra[#1 #2 #3.#4]{#1\sup{h}#2\sup{m}#3\sup{s}\llap.#4}
\def\dec[#1 #2 #3.#4]{#1\deg#2\arcm#3\arcs\llap.#4}
\def\deco[#1 #2 #3]{#1\deg#2\arcm#3\arcs}
\def\rra[#1 #2]{#1\sup{h}#2\sup{m}}

\def\dots{\relax\ifmmode \ldots\else $\ldots$\fi}
\def\WHzsr{\ifmmode $W\,Hz\mo\,sr\mo$\else W\,Hz\mo\,sr\mo\fi}
\def\mHz{\ifmmode $\,mHz$\else \,mHz\fi}
\def\GHz{\ifmmode $\,GHz$\else \,GHz\fi}
\def\mKs{\ifmmode $\,mK\,s$^{1/2}\else \,mK\,s$^{1/2}$\fi}
\def\muKs{\ifmmode \,\mu$K\,s$^{1/2}\else \,$\mu$K\,s$^{1/2}$\fi}
\def\muKRJs{\ifmmode \,\mu$K$_{\rm RJ}$\,s$^{1/2}\else \,$\mu$K$_{\rm RJ}$\,s$^{1/2}$\fi}
\def\muKHz{\ifmmode \,\mu$K\,Hz$^{-1/2}\else \,$\mu$K\,Hz$^{-1/2}$\fi}
\def\MJysr{\ifmmode \,$MJy\,sr\mo$\else \,MJy\,sr\mo\fi}
\def\MJysrmK{\ifmmode \,$MJy\,sr\mo$\,mK$_{\rm CMB}\mo\else \,MJy\,sr\mo\,mK$_{\rm CMB}\mo$\fi}
\def\microns{\ifmmode \,\mu$m$\else \,$\mu$m\fi}

\def\muK{\ifmmode \,\mu$K$\else \,$\mu$\hbox{K}\fi}
\def\microK{\ifmmode \,\mu$K$\else \,$\mu$\hbox{K}\fi}
\def\muW{\ifmmode \,\mu$W$\else \,$\mu$\hbox{W}\fi}
\def\kms{\ifmmode $\,km\,s$^{-1}\else \,km\,s$^{-1}$\fi}
\def\kmsMpc{\ifmmode $\,\kms\,Mpc\mo$\else \,\kms\,Mpc\mo\fi}
\setsymbol{LFI:center:frequency:70GHz:units}{70.3\,GHz}
\setsymbol{LFI:center:frequency:44GHz:units}{44.1\,GHz}
\setsymbol{LFI:center:frequency:30GHz:units}{28.5\,GHz}

\setsymbol{LFI:center:frequency:70GHz}{70.3}
\setsymbol{LFI:center:frequency:44GHz}{44.1}
\setsymbol{LFI:center:frequency:30GHz}{28.5}

\setsymbol{LFI:center:frequency:LFI18:Rad:M:units}{71.7\GHz}
\setsymbol{LFI:center:frequency:LFI19:Rad:M:units}{67.5\GHz}
\setsymbol{LFI:center:frequency:LFI20:Rad:M:units}{69.2\GHz}
\setsymbol{LFI:center:frequency:LFI21:Rad:M:units}{70.4\GHz}
\setsymbol{LFI:center:frequency:LFI22:Rad:M:units}{71.5\GHz}
\setsymbol{LFI:center:frequency:LFI23:Rad:M:units}{70.8\GHz}
\setsymbol{LFI:center:frequency:LFI24:Rad:M:units}{44.4\GHz}
\setsymbol{LFI:center:frequency:LFI25:Rad:M:units}{44.0\GHz}
\setsymbol{LFI:center:frequency:LFI26:Rad:M:units}{43.9\GHz}
\setsymbol{LFI:center:frequency:LFI27:Rad:M:units}{28.3\GHz}
\setsymbol{LFI:center:frequency:LFI28:Rad:M:units}{28.8\GHz}
\setsymbol{LFI:center:frequency:LFI18:Rad:S:units}{70.1\GHz}
\setsymbol{LFI:center:frequency:LFI19:Rad:S:units}{69.6\GHz}
\setsymbol{LFI:center:frequency:LFI20:Rad:S:units}{69.5\GHz}
\setsymbol{LFI:center:frequency:LFI21:Rad:S:units}{69.5\GHz}
\setsymbol{LFI:center:frequency:LFI22:Rad:S:units}{72.8\GHz}
\setsymbol{LFI:center:frequency:LFI23:Rad:S:units}{71.3\GHz}
\setsymbol{LFI:center:frequency:LFI24:Rad:S:units}{44.1\GHz}
\setsymbol{LFI:center:frequency:LFI25:Rad:S:units}{44.1\GHz}
\setsymbol{LFI:center:frequency:LFI26:Rad:S:units}{44.1\GHz}
\setsymbol{LFI:center:frequency:LFI27:Rad:S:units}{28.5\GHz}
\setsymbol{LFI:center:frequency:LFI28:Rad:S:units}{28.2\GHz}

\setsymbol{LFI:center:frequency:LFI18:Rad:M}{71.7}
\setsymbol{LFI:center:frequency:LFI19:Rad:M}{67.5}
\setsymbol{LFI:center:frequency:LFI20:Rad:M}{69.2}
\setsymbol{LFI:center:frequency:LFI21:Rad:M}{70.4}
\setsymbol{LFI:center:frequency:LFI22:Rad:M}{71.5}
\setsymbol{LFI:center:frequency:LFI23:Rad:M}{70.8}
\setsymbol{LFI:center:frequency:LFI24:Rad:M}{44.4}
\setsymbol{LFI:center:frequency:LFI25:Rad:M}{44.0}
\setsymbol{LFI:center:frequency:LFI26:Rad:M}{43.9}
\setsymbol{LFI:center:frequency:LFI27:Rad:M}{28.3}
\setsymbol{LFI:center:frequency:LFI28:Rad:M}{28.8}
\setsymbol{LFI:center:frequency:LFI18:Rad:S}{70.1}
\setsymbol{LFI:center:frequency:LFI19:Rad:S}{69.6}
\setsymbol{LFI:center:frequency:LFI20:Rad:S}{69.5}
\setsymbol{LFI:center:frequency:LFI21:Rad:S}{69.5}
\setsymbol{LFI:center:frequency:LFI22:Rad:S}{72.8}
\setsymbol{LFI:center:frequency:LFI23:Rad:S}{71.3}
\setsymbol{LFI:center:frequency:LFI24:Rad:S}{44.1}
\setsymbol{LFI:center:frequency:LFI25:Rad:S}{44.1}
\setsymbol{LFI:center:frequency:LFI26:Rad:S}{44.1}
\setsymbol{LFI:center:frequency:LFI27:Rad:S}{28.5}
\setsymbol{LFI:center:frequency:LFI28:Rad:S}{28.2}

\setsymbol{LFI:white:noise:sensitivity:70GHz:units}{134.7\muKs}
\setsymbol{LFI:white:noise:sensitivity:44GHz:units}{164.7\muKs}
\setsymbol{LFI:white:noise:sensitivity:30GHz:units}{143.4\muKs}

\setsymbol{LFI:white:noise:sensitivity:70GHz}{134.7}
\setsymbol{LFI:white:noise:sensitivity:44GHz}{164.7}
\setsymbol{LFI:white:noise:sensitivity:30GHz}{143.4}

\setsymbol{LFI:white:noise:sensitivity:LFI18:Rad:M:units}{512.0\muKs}
\setsymbol{LFI:white:noise:sensitivity:LFI19:Rad:M:units}{581.4\muKs}
\setsymbol{LFI:white:noise:sensitivity:LFI20:Rad:M:units}{590.8\muKs}
\setsymbol{LFI:white:noise:sensitivity:LFI21:Rad:M:units}{455.2\muKs}
\setsymbol{LFI:white:noise:sensitivity:LFI22:Rad:M:units}{492.0\muKs}
\setsymbol{LFI:white:noise:sensitivity:LFI23:Rad:M:units}{507.7\muKs}
\setsymbol{LFI:white:noise:sensitivity:LFI24:Rad:M:units}{462.2\muKs}
\setsymbol{LFI:white:noise:sensitivity:LFI25:Rad:M:units}{413.6\muKs}
\setsymbol{LFI:white:noise:sensitivity:LFI26:Rad:M:units}{478.6\muKs}
\setsymbol{LFI:white:noise:sensitivity:LFI27:Rad:M:units}{277.7\muKs}
\setsymbol{LFI:white:noise:sensitivity:LFI28:Rad:M:units}{312.3\muKs}
\setsymbol{LFI:white:noise:sensitivity:LFI18:Rad:S:units}{465.7\muKs}
\setsymbol{LFI:white:noise:sensitivity:LFI19:Rad:S:units}{555.6\muKs}
\setsymbol{LFI:white:noise:sensitivity:LFI20:Rad:S:units}{623.2\muKs}
\setsymbol{LFI:white:noise:sensitivity:LFI21:Rad:S:units}{564.1\muKs}
\setsymbol{LFI:white:noise:sensitivity:LFI22:Rad:S:units}{534.4\muKs}
\setsymbol{LFI:white:noise:sensitivity:LFI23:Rad:S:units}{542.4\muKs}
\setsymbol{LFI:white:noise:sensitivity:LFI24:Rad:S:units}{399.2\muKs}
\setsymbol{LFI:white:noise:sensitivity:LFI25:Rad:S:units}{392.6\muKs}
\setsymbol{LFI:white:noise:sensitivity:LFI26:Rad:S:units}{418.6\muKs}
\setsymbol{LFI:white:noise:sensitivity:LFI27:Rad:S:units}{302.9\muKs}
\setsymbol{LFI:white:noise:sensitivity:LFI28:Rad:S:units}{285.3\muKs}

\setsymbol{LFI:white:noise:sensitivity:LFI18:Rad:M}{512.0}
\setsymbol{LFI:white:noise:sensitivity:LFI19:Rad:M}{581.4}
\setsymbol{LFI:white:noise:sensitivity:LFI20:Rad:M}{590.8}
\setsymbol{LFI:white:noise:sensitivity:LFI21:Rad:M}{455.2}
\setsymbol{LFI:white:noise:sensitivity:LFI22:Rad:M}{492.0}
\setsymbol{LFI:white:noise:sensitivity:LFI23:Rad:M}{507.7}
\setsymbol{LFI:white:noise:sensitivity:LFI24:Rad:M}{462.2}
\setsymbol{LFI:white:noise:sensitivity:LFI25:Rad:M}{413.6}
\setsymbol{LFI:white:noise:sensitivity:LFI26:Rad:M}{478.6}
\setsymbol{LFI:white:noise:sensitivity:LFI27:Rad:M}{277.7}
\setsymbol{LFI:white:noise:sensitivity:LFI28:Rad:M}{312.3}
\setsymbol{LFI:white:noise:sensitivity:LFI18:Rad:S}{465.7}
\setsymbol{LFI:white:noise:sensitivity:LFI19:Rad:S}{555.6}
\setsymbol{LFI:white:noise:sensitivity:LFI20:Rad:S}{623.2}
\setsymbol{LFI:white:noise:sensitivity:LFI21:Rad:S}{564.1}
\setsymbol{LFI:white:noise:sensitivity:LFI22:Rad:S}{534.4}
\setsymbol{LFI:white:noise:sensitivity:LFI23:Rad:S}{542.4}
\setsymbol{LFI:white:noise:sensitivity:LFI24:Rad:S}{399.2}
\setsymbol{LFI:white:noise:sensitivity:LFI25:Rad:S}{392.6}
\setsymbol{LFI:white:noise:sensitivity:LFI26:Rad:S}{418.6}
\setsymbol{LFI:white:noise:sensitivity:LFI27:Rad:S}{302.9}
\setsymbol{LFI:white:noise:sensitivity:LFI28:Rad:S}{285.3}

\setsymbol{LFI:knee:frequency:70GHz:units}{29.5\mHz}
\setsymbol{LFI:knee:frequency:44GHz:units}{56.2\mHz}
\setsymbol{LFI:knee:frequency:30GHz:units}{113.7\mHz}

\setsymbol{LFI:knee:frequency:70GHz}{29.5}
\setsymbol{LFI:knee:frequency:44GHz}{56.2}
\setsymbol{LFI:knee:frequency:30GHz}{113.7}

\setsymbol{LFI:knee:frequency:LFI18:Rad:M:units}{16.3\mHz}
\setsymbol{LFI:knee:frequency:LFI19:Rad:M:units}{15.1\mHz}
\setsymbol{LFI:knee:frequency:LFI20:Rad:M:units}{18.7\mHz}
\setsymbol{LFI:knee:frequency:LFI21:Rad:M:units}{37.2\mHz}
\setsymbol{LFI:knee:frequency:LFI22:Rad:M:units}{12.7\mHz}
\setsymbol{LFI:knee:frequency:LFI23:Rad:M:units}{34.6\mHz}
\setsymbol{LFI:knee:frequency:LFI24:Rad:M:units}{46.2\mHz}
\setsymbol{LFI:knee:frequency:LFI25:Rad:M:units}{24.9\mHz}
\setsymbol{LFI:knee:frequency:LFI26:Rad:M:units}{67.6\mHz}
\setsymbol{LFI:knee:frequency:LFI27:Rad:M:units}{187.4\mHz}
\setsymbol{LFI:knee:frequency:LFI28:Rad:M:units}{122.2\mHz}
\setsymbol{LFI:knee:frequency:LFI18:Rad:S:units}{17.7\mHz}
\setsymbol{LFI:knee:frequency:LFI19:Rad:S:units}{22.0\mHz}
\setsymbol{LFI:knee:frequency:LFI20:Rad:S:units}{8.7\mHz}
\setsymbol{LFI:knee:frequency:LFI21:Rad:S:units}{25.9\mHz}
\setsymbol{LFI:knee:frequency:LFI22:Rad:S:units}{15.8\mHz}
\setsymbol{LFI:knee:frequency:LFI23:Rad:S:units}{129.8\mHz}
\setsymbol{LFI:knee:frequency:LFI24:Rad:S:units}{100.9\mHz}
\setsymbol{LFI:knee:frequency:LFI25:Rad:S:units}{38.9\mHz}
\setsymbol{LFI:knee:frequency:LFI26:Rad:S:units}{58.9\mHz}
\setsymbol{LFI:knee:frequency:LFI27:Rad:S:units}{104.4\mHz}
\setsymbol{LFI:knee:frequency:LFI28:Rad:S:units}{40.7\mHz}

\setsymbol{LFI:knee:frequency:LFI18:Rad:M}{16.3}
\setsymbol{LFI:knee:frequency:LFI19:Rad:M}{15.1}
\setsymbol{LFI:knee:frequency:LFI20:Rad:M}{18.7}
\setsymbol{LFI:knee:frequency:LFI21:Rad:M}{37.2}
\setsymbol{LFI:knee:frequency:LFI22:Rad:M}{12.7}
\setsymbol{LFI:knee:frequency:LFI23:Rad:M}{34.6}
\setsymbol{LFI:knee:frequency:LFI24:Rad:M}{46.2}
\setsymbol{LFI:knee:frequency:LFI25:Rad:M}{24.9}
\setsymbol{LFI:knee:frequency:LFI26:Rad:M}{67.6}
\setsymbol{LFI:knee:frequency:LFI27:Rad:M}{187.4}
\setsymbol{LFI:knee:frequency:LFI28:Rad:M}{122.2}
\setsymbol{LFI:knee:frequency:LFI18:Rad:S}{17.7}
\setsymbol{LFI:knee:frequency:LFI19:Rad:S}{22.0}
\setsymbol{LFI:knee:frequency:LFI20:Rad:S}{8.7}
\setsymbol{LFI:knee:frequency:LFI21:Rad:S}{25.9}
\setsymbol{LFI:knee:frequency:LFI22:Rad:S}{15.8}
\setsymbol{LFI:knee:frequency:LFI23:Rad:S}{129.8}
\setsymbol{LFI:knee:frequency:LFI24:Rad:S}{100.9}
\setsymbol{LFI:knee:frequency:LFI25:Rad:S}{38.9}
\setsymbol{LFI:knee:frequency:LFI26:Rad:S}{58.9}
\setsymbol{LFI:knee:frequency:LFI27:Rad:S}{104.4}
\setsymbol{LFI:knee:frequency:LFI28:Rad:S}{40.7}

\setsymbol{LFI:slope:70GHz:units}{$-1.03$\mHz}
\setsymbol{LFI:slope:44GHz:units}{$-0.89$\mHz}
\setsymbol{LFI:slope:30GHz:units}{$-0.87$\mHz}

\setsymbol{LFI:slope:70GHz}{$-1.03$}
\setsymbol{LFI:slope:44GHz}{$-0.89$}
\setsymbol{LFI:slope:30GHz}{$-0.87$}

\setsymbol{LFI:slope:LFI18:Rad:M:units}{$-1.04$\mHz}
\setsymbol{LFI:slope:LFI19:Rad:M:units}{$-1.09$\mHz}
\setsymbol{LFI:slope:LFI20:Rad:M:units}{$-0.69$\mHz}
\setsymbol{LFI:slope:LFI21:Rad:M:units}{$-1.56$\mHz}
\setsymbol{LFI:slope:LFI22:Rad:M:units}{$-1.01$\mHz}
\setsymbol{LFI:slope:LFI23:Rad:M:units}{$-0.96$\mHz}
\setsymbol{LFI:slope:LFI24:Rad:M:units}{$-0.83$\mHz}
\setsymbol{LFI:slope:LFI25:Rad:M:units}{$-0.91$\mHz}
\setsymbol{LFI:slope:LFI26:Rad:M:units}{$-0.95$\mHz}
\setsymbol{LFI:slope:LFI27:Rad:M:units}{$-0.87$\mHz}
\setsymbol{LFI:slope:LFI28:Rad:M:units}{$-0.88$\mHz}
\setsymbol{LFI:slope:LFI18:Rad:S:units}{$-1.15$\mHz}
\setsymbol{LFI:slope:LFI19:Rad:S:units}{$-1.00$\mHz}
\setsymbol{LFI:slope:LFI20:Rad:S:units}{$-0.95$\mHz}
\setsymbol{LFI:slope:LFI21:Rad:S:units}{$-0.92$\mHz}
\setsymbol{LFI:slope:LFI22:Rad:S:units}{$-1.01$\mHz}
\setsymbol{LFI:slope:LFI23:Rad:S:units}{$-0.95$\mHz}
\setsymbol{LFI:slope:LFI24:Rad:S:units}{$-0.73$\mHz}
\setsymbol{LFI:slope:LFI25:Rad:S:units}{$-1.16$\mHz}
\setsymbol{LFI:slope:LFI26:Rad:S:units}{$-0.79$\mHz}
\setsymbol{LFI:slope:LFI27:Rad:S:units}{$-0.82$\mHz}
\setsymbol{LFI:slope:LFI28:Rad:S:units}{$-0.91$\mHz}

\setsymbol{LFI:slope:LFI18:Rad:M}{$-1.04$}
\setsymbol{LFI:slope:LFI19:Rad:M}{$-1.09$}
\setsymbol{LFI:slope:LFI20:Rad:M}{$-0.69$}
\setsymbol{LFI:slope:LFI21:Rad:M}{$-1.56$}
\setsymbol{LFI:slope:LFI22:Rad:M}{$-1.01$}
\setsymbol{LFI:slope:LFI23:Rad:M}{$-0.96$}
\setsymbol{LFI:slope:LFI24:Rad:M}{$-0.83$}
\setsymbol{LFI:slope:LFI25:Rad:M}{$-0.91$}
\setsymbol{LFI:slope:LFI26:Rad:M}{$-0.95$}
\setsymbol{LFI:slope:LFI27:Rad:M}{$-0.87$}
\setsymbol{LFI:slope:LFI28:Rad:M}{$-0.88$}
\setsymbol{LFI:slope:LFI18:Rad:S}{$-1.15$}
\setsymbol{LFI:slope:LFI19:Rad:S}{$-1.00$}
\setsymbol{LFI:slope:LFI20:Rad:S}{$-0.95$}
\setsymbol{LFI:slope:LFI21:Rad:S}{$-0.92$}
\setsymbol{LFI:slope:LFI22:Rad:S}{$-1.01$}
\setsymbol{LFI:slope:LFI23:Rad:S}{$-0.95$}
\setsymbol{LFI:slope:LFI24:Rad:S}{$-0.73$}
\setsymbol{LFI:slope:LFI25:Rad:S}{$-1.16$}
\setsymbol{LFI:slope:LFI26:Rad:S}{$-0.79$}
\setsymbol{LFI:slope:LFI27:Rad:S}{$-0.82$}
\setsymbol{LFI:slope:LFI28:Rad:S}{$-0.91$}

\setsymbol{LFI:FWHM:70GHz:units}{13\parcm01}
\setsymbol{LFI:FWHM:44GHz:units}{27\parcm92}
\setsymbol{LFI:FWHM:30GHz:units}{32\parcm65}

\setsymbol{LFI:FWHM:70GHz}{13.01}
\setsymbol{LFI:FWHM:44GHz}{27.92}
\setsymbol{LFI:FWHM:30GHz}{32.65}

\setsymbol{LFI:FWHM:LFI18:units}{13\parcm39}
\setsymbol{LFI:FWHM:LFI19:units}{13\parcm01}
\setsymbol{LFI:FWHM:LFI20:units}{12\parcm75}
\setsymbol{LFI:FWHM:LFI21:units}{12\parcm74}
\setsymbol{LFI:FWHM:LFI22:units}{12\parcm87}
\setsymbol{LFI:FWHM:LFI23:units}{13\parcm27}
\setsymbol{LFI:FWHM:LFI24:units}{22\parcm98}
\setsymbol{LFI:FWHM:LFI25:units}{30\parcm46}
\setsymbol{LFI:FWHM:LFI26:units}{30\parcm31}
\setsymbol{LFI:FWHM:LFI27:units}{32\parcm65}
\setsymbol{LFI:FWHM:LFI28:units}{32\parcm66}

\setsymbol{LFI:FWHM:LFI18}{13.39}
\setsymbol{LFI:FWHM:LFI19}{13.01}
\setsymbol{LFI:FWHM:LFI20}{12.75}
\setsymbol{LFI:FWHM:LFI21}{12.74}
\setsymbol{LFI:FWHM:LFI22}{12.87}
\setsymbol{LFI:FWHM:LFI23}{13.27}
\setsymbol{LFI:FWHM:LFI24}{22.98}
\setsymbol{LFI:FWHM:LFI25}{30.46}
\setsymbol{LFI:FWHM:LFI26}{30.31}
\setsymbol{LFI:FWHM:LFI27}{32.65}
\setsymbol{LFI:FWHM:LFI28}{32.66}

\setsymbol{LFI:FWHM:uncertainty:LFI18:units}{0.170\arcm}
\setsymbol{LFI:FWHM:uncertainty:LFI19:units}{0.174\arcm}
\setsymbol{LFI:FWHM:uncertainty:LFI20:units}{0.170\arcm}
\setsymbol{LFI:FWHM:uncertainty:LFI21:units}{0.156\arcm}
\setsymbol{LFI:FWHM:uncertainty:LFI22:units}{0.164\arcm}
\setsymbol{LFI:FWHM:uncertainty:LFI23:units}{0.171\arcm}
\setsymbol{LFI:FWHM:uncertainty:LFI24:units}{0.652\arcm}
\setsymbol{LFI:FWHM:uncertainty:LFI25:units}{1.075\arcm}
\setsymbol{LFI:FWHM:uncertainty:LFI26:units}{1.131\arcm}
\setsymbol{LFI:FWHM:uncertainty:LFI27:units}{1.266\arcm}
\setsymbol{LFI:FWHM:uncertainty:LFI28:units}{1.287\arcm}

\setsymbol{LFI:FWHM:uncertainty:LFI18}{0.170}
\setsymbol{LFI:FWHM:uncertainty:LFI19}{0.174}
\setsymbol{LFI:FWHM:uncertainty:LFI20}{0.170}
\setsymbol{LFI:FWHM:uncertainty:LFI21}{0.156}
\setsymbol{LFI:FWHM:uncertainty:LFI22}{0.164}
\setsymbol{LFI:FWHM:uncertainty:LFI23}{0.171}
\setsymbol{LFI:FWHM:uncertainty:LFI24}{0.652}
\setsymbol{LFI:FWHM:uncertainty:LFI25}{1.075}
\setsymbol{LFI:FWHM:uncertainty:LFI26}{1.131}
\setsymbol{LFI:FWHM:uncertainty:LFI27}{1.266}
\setsymbol{LFI:FWHM:uncertainty:LFI28}{1.287}

\setsymbol{HFI:center:frequency:100GHz:units}{100\,GHz}
\setsymbol{HFI:center:frequency:143GHz:units}{143\,GHz}
\setsymbol{HFI:center:frequency:217GHz:units}{217\,GHz}
\setsymbol{HFI:center:frequency:353GHz:units}{353\,GHz}
\setsymbol{HFI:center:frequency:545GHz:units}{545\,GHz}
\setsymbol{HFI:center:frequency:857GHz:units}{857\,GHz}

\setsymbol{HFI:center:frequency:100GHz}{100}
\setsymbol{HFI:center:frequency:143GHz}{143}
\setsymbol{HFI:center:frequency:217GHz}{217}
\setsymbol{HFI:center:frequency:353GHz}{353}
\setsymbol{HFI:center:frequency:545GHz}{545}
\setsymbol{HFI:center:frequency:857GHz}{857}

\setsymbol{HFI:Ndetectors:100GHz}{8}
\setsymbol{HFI:Ndetectors:143GHz}{11}
\setsymbol{HFI:Ndetectors:217GHz}{12}
\setsymbol{HFI:Ndetectors:353GHz}{12}
\setsymbol{HFI:Ndetectors:545GHz}{3}
\setsymbol{HFI:Ndetectors:857GHz}{4}

\setsymbol{HFI:FWHM:Maps:100GHz:units}{9\parcm88}
\setsymbol{HFI:FWHM:Maps:143GHz:units}{7\parcm18}
\setsymbol{HFI:FWHM:Maps:217GHz:units}{4\parcm87}
\setsymbol{HFI:FWHM:Maps:353GHz:units}{4\parcm65}
\setsymbol{HFI:FWHM:Maps:545GHz:units}{4\parcm72}
\setsymbol{HFI:FWHM:Maps:857GHz:units}{4\parcm39}
\setsymbol{HFI:FWHM:Maps:100GHz}{9.88}
\setsymbol{HFI:FWHM:Maps:143GHz}{7.18}
\setsymbol{HFI:FWHM:Maps:217GHz}{4.87}
\setsymbol{HFI:FWHM:Maps:353GHz}{4.65}
\setsymbol{HFI:FWHM:Maps:545GHz}{4.72}
\setsymbol{HFI:FWHM:Maps:857GHz}{4.39}

\setsymbol{HFI:beam:ellipticity:Maps:100GHz}{1.15}
\setsymbol{HFI:beam:ellipticity:Maps:143GHz}{1.01}
\setsymbol{HFI:beam:ellipticity:Maps:217GHz}{1.06}
\setsymbol{HFI:beam:ellipticity:Maps:353GHz}{1.05}
\setsymbol{HFI:beam:ellipticity:Maps:545GHz}{1.14}
\setsymbol{HFI:beam:ellipticity:Maps:857GHz}{1.19}

\setsymbol{HFI:FWHM:Mars:100GHz:units}{9\parcm37}
\setsymbol{HFI:FWHM:Mars:143GHz:units}{7\parcm04}
\setsymbol{HFI:FWHM:Mars:217GHz:units}{4\parcm68}
\setsymbol{HFI:FWHM:Mars:353GHz:units}{4\parcm43}
\setsymbol{HFI:FWHM:Mars:545GHz:units}{3\parcm80}
\setsymbol{HFI:FWHM:Mars:857GHz:units}{3\parcm67}

\setsymbol{HFI:FWHM:Mars:100GHz}{9.37}
\setsymbol{HFI:FWHM:Mars:143GHz}{7.04}
\setsymbol{HFI:FWHM:Mars:217GHz}{4.68}
\setsymbol{HFI:FWHM:Mars:353GHz}{4.43}
\setsymbol{HFI:FWHM:Mars:545GHz}{3.80}
\setsymbol{HFI:FWHM:Mars:857GHz}{3.67}

\setsymbol{HFI:beam:ellipticity:Mars:100GHz}{1.18}
\setsymbol{HFI:beam:ellipticity:Mars:143GHz}{1.03}
\setsymbol{HFI:beam:ellipticity:Mars:217GHz}{1.14}
\setsymbol{HFI:beam:ellipticity:Mars:353GHz}{1.09}
\setsymbol{HFI:beam:ellipticity:Mars:545GHz}{1.25}
\setsymbol{HFI:beam:ellipticity:Mars:857GHz}{1.03}

\setsymbol{HFI:CMB:relative:calibration:100GHz}{$\lsim 1\%$}
\setsymbol{HFI:CMB:relative:calibration:143GHz}{$\lsim 1\%$}
\setsymbol{HFI:CMB:relative:calibration:217GHz}{$\lsim 1\%$}
\setsymbol{HFI:CMB:relative:calibration:353GHz}{$\lsim 1\%$}
\setsymbol{HFI:CMB:relative:calibration:545GHz}{}
\setsymbol{HFI:CMB:relative:calibration:857GHz}{}

\setsymbol{HFI:CMB:absolute:calibration:100GHz}{$\lsim 2\%$}
\setsymbol{HFI:CMB:absolute:calibration:143GHz}{$\lsim 2\%$}
\setsymbol{HFI:CMB:absolute:calibration:217GHz}{$\lsim 2\%$}
\setsymbol{HFI:CMB:absolute:calibration:353GHz}{$\lsim 2\%$}
\setsymbol{HFI:CMB:absolute:calibration:545GHz}{}
\setsymbol{HFI:CMB:absolute:calibration:857GHz}{}

\setsymbol{HFI:FIRAS:gain:calibration:accuracy:statistical:100GHz}{}
\setsymbol{HFI:FIRAS:gain:calibration:accuracy:statistical:143GHz}{}
\setsymbol{HFI:FIRAS:gain:calibration:accuracy:statistical:217GHz}{}
\setsymbol{HFI:FIRAS:gain:calibration:accuracy:statistical:353GHz}{2.5\%}
\setsymbol{HFI:FIRAS:gain:calibration:accuracy:statistical:545GHz}{1\%}
\setsymbol{HFI:FIRAS:gain:calibration:accuracy:statistical:857GHz}{0.5\%}

\setsymbol{HFI:FIRAS:gain:calibration:accuracy:systematic:100GHz}{}
\setsymbol{HFI:FIRAS:gain:calibration:accuracy:systematic:143GHz}{}
\setsymbol{HFI:FIRAS:gain:calibration:accuracy:systematic:217GHz}{}
\setsymbol{HFI:FIRAS:gain:calibration:accuracy:systematic:353GHz}{}
\setsymbol{HFI:FIRAS:gain:calibration:accuracy:systematic:545GHz}{7\%}
\setsymbol{HFI:FIRAS:gain:calibration:accuracy:systematic:857GHz}{7\%}

\setsymbol{HFI:FIRAS:zero:point:accuracy:100GHz:units}{0.8\MJysr}
\setsymbol{HFI:FIRAS:zero:point:accuracy:143GHz:units}{}
\setsymbol{HFI:FIRAS:zero:point:accuracy:217GHz:units}{}
\setsymbol{HFI:FIRAS:zero:point:accuracy:353GHz:units}{1.4\MJysr}
\setsymbol{HFI:FIRAS:zero:point:accuracy:545GHz:units}{2.2\MJysr}
\setsymbol{HFI:FIRAS:zero:point:accuracy:857GHz:units}{1.7\MJysr}

\setsymbol{HFI:FIRAS:zero:point:accuracy:100GHz}{0.8}
\setsymbol{HFI:FIRAS:zero:point:accuracy:143GHz}{}
\setsymbol{HFI:FIRAS:zero:point:accuracy:217GHz}{}
\setsymbol{HFI:FIRAS:zero:point:accuracy:353GHz}{1.4}
\setsymbol{HFI:FIRAS:zero:point:accuracy:545GHz}{2.2}
\setsymbol{HFI:FIRAS:zero:point:accuracy:857GHz}{1.7}

\setsymbol{HFI:unit:conversion:100GHz:units}{0.2415\MJysrmK}
\setsymbol{HFI:unit:conversion:143GHz:units}{0.3694\MJysrmK}
\setsymbol{HFI:unit:conversion:217GHz:units}{0.4811\MJysrmK}
\setsymbol{HFI:unit:conversion:353GHz:units}{0.2883\MJysrmK}
\setsymbol{HFI:unit:conversion:545GHz:units}{0.05826\MJysrmK}
\setsymbol{HFI:unit:conversion:857GHz:units}{0.002238\MJysrmK}

\setsymbol{HFI:unit:conversion:100GHz}{0.2415}
\setsymbol{HFI:unit:conversion:143GHz}{0.3694}
\setsymbol{HFI:unit:conversion:217GHz}{0.4811}
\setsymbol{HFI:unit:conversion:353GHz}{0.2883}
\setsymbol{HFI:unit:conversion:545GHz}{0.05826}
\setsymbol{HFI:unit:conversion:857GHz}{0.002238}

\setsymbol{HFI:colour:correction:alpha=-2:V1.01:100GHz}{0.9893}
\setsymbol{HFI:colour:correction:alpha=-2:V1.01:143GHz}{0.9759}
\setsymbol{HFI:colour:correction:alpha=-2:V1.01:217GHz}{1.0007}
\setsymbol{HFI:colour:correction:alpha=-2:V1.01:353GHz}{1.0028}
\setsymbol{HFI:colour:correction:alpha=-2:V1.01:545GHz}{1.0019}
\setsymbol{HFI:colour:correction:alpha=-2:V1.01:857GHz}{0.9889}

\setsymbol{HFI:colour:correction:alpha=0:V1.01:100GHz}{1.0008}
\setsymbol{HFI:colour:correction:alpha=0:V1.01:143GHz}{1.0148}
\setsymbol{HFI:colour:correction:alpha=0:V1.01:217GHz}{0.9909}
\setsymbol{HFI:colour:correction:alpha=0:V1.01:353GHz}{0.9888}
\setsymbol{HFI:colour:correction:alpha=0:V1.01:545GHz}{0.9878}
\setsymbol{HFI:colour:correction:alpha=0:V1.01:857GHz}{1.0014}

\begin{document}

\title{\textit{Planck} Intermediate Results. XI: The gas content of dark matter halos:
  the Sunyaev-Zeldovich-stellar mass relation for locally brightest galaxies}

\titlerunning{Gas content of dark matter halos}

\author{\small
Planck Collaboration:
P.~A.~R.~Ade\inst{80}
\and
N.~Aghanim\inst{55}
\and
M.~Arnaud\inst{69}
\and
M.~Ashdown\inst{66, 7}
\and
F.~Atrio-Barandela\inst{20}
\and
J.~Aumont\inst{55}
\and
C.~Baccigalupi\inst{79}
\and
A.~J.~Banday\inst{89, 10}
\and
R.~B.~Barreiro\inst{63}
\and
R.~Barrena\inst{62}
% ...
}
\institute{\small
APC, AstroParticule et Cosmologie, Universit\'{e} Paris Diderot, CNRS/IN2P3, CEA/lrfu, Observatoire de Paris, Sorbonne Paris Cit\'{e}, 10, rue Alice Domon et L\'{e}onie Duquet, 75205 Paris Cedex 13, France\\
\and
Aalto University Mets\"{a}hovi Radio Observatory, Mets\"{a}hovintie 114, FIN-02540 Kylm\"{a}l\"{a}, Finland\\
\and
Academy of Sciences of Tatarstan, Bauman Str., 20, Kazan, 420111, Republic of Tatarstan, Russia\\
\and
African Institute for Mathematical Sciences, 6-8 Melrose Road, Muizenberg, Cape Town, South Africa\\
\and
Agenzia Spaziale Italiana Science Data Center, c/o ESRIN, via Galileo Galilei, Frascati, Italy\\
\and
Agenzia Spaziale Italiana, Viale Liegi 26, Roma, Italy\\
\and
Astrophysics Group, Cavendish Laboratory, University of Cambridge, J J Thomson Avenue, Cambridge CB3 0HE, U.K.\\
\and
Atacama Large Millimeter/submillimeter Array, ALMA Santiago Central Offices, Alonso de Cordova 3107, Vitacura, Casilla 763 0355, Santiago, Chile\\
\and
CITA, University of Toronto, 60 St. George St., Toronto, ON M5S 3H8, Canada\\
\and
CNRS, IRAP, 9 Av. colonel Roche, BP 44346, F-31028 Toulouse cedex 4, France\\
\and
California Institute of Technology, Pasadena, California, U.S.A.\\
\and
Centre of Mathematics for Applications, University of Oslo, Blindern, Oslo, Norway\\
\and
Centro de Astrof\'{\i}sica, Universidade do Porto, Rua das Estrelas, 4150-762 Porto, Portugal\\
\and
Centro de Estudios de F\'{i}sica del Cosmos de Arag\'{o}n (CEFCA), Plaza San Juan, 1, planta 2, E-44001, Teruel, Spain\\
\and
Computational Cosmology Center, Lawrence Berkeley National Laboratory, Berkeley, California, U.S.A.\\
\and
Consejo Superior de Investigaciones Cient\'{\i}ficas (CSIC), Madrid, Spain\\
\and
DSM/Irfu/SPP, CEA-Saclay, F-91191 Gif-sur-Yvette Cedex, France\\
\and
DTU Space, National Space Institute, Technical University of Denmark, Elektrovej 327, DK-2800 Kgs. Lyngby, Denmark\\
\and
D\'{e}partement de Physique Th\'{e}orique, Universit\'{e} de Gen\`{e}ve, 24, Quai E. Ansermet,1211 Gen\`{e}ve 4, Switzerland\\
\and
Departamento de F\'{\i}sica Fundamental, Facultad de Ciencias, Universidad de Salamanca, 37008 Salamanca, Spain\\
\and
Departamento de F\'{\i}sica, Universidad de Oviedo, Avda. Calvo Sotelo s/n, Oviedo, Spain\\
\and
Department of Astronomy and Geodesy, Kazan Federal University,  Kremlevskaya Str., 18, Kazan, 420008, Russia\\
\and
Department of Astrophysics/IMAPP, Radboud University Nijmegen, P.O. Box 9010, 6500 GL Nijmegen, The Netherlands\\
\and
Department of Physics \& Astronomy, University of British Columbia, 6224 Agricultural Road, Vancouver, British Columbia, Canada\\
\and
Department of Physics and Astronomy, Dana and David Dornsife College of Letter, Arts and Sciences, University of Southern California, Los Angeles, CA 90089, U.S.A.\\
\and
Department of Physics, Gustaf H\"{a}llstr\"{o}min katu 2a, University of Helsinki, Helsinki, Finland\\
\and
Department of Physics, Princeton University, Princeton, New Jersey, U.S.A.\\
\and
Department of Physics, University of California, Berkeley, California, U.S.A.\\
\and
Department of Physics, University of California, Santa Barbara, California, U.S.A.\\
\and
Department of Physics, University of Illinois at Urbana-Champaign, 1110 West Green Street, Urbana, Illinois, U.S.A.\\
\and
Dipartimento di Fisica e Astronomia G. Galilei, Universit\`{a} degli Studi di Padova, via Marzolo 8, 35131 Padova, Italy\\
\and
Dipartimento di Fisica e Scienze della Terra, Universit\`{a} di Ferrara, Via Saragat 1, 44122 Ferrara, Italy\\
\and
Dipartimento di Fisica, Universit\`{a} La Sapienza, P. le A. Moro 2, Roma, Italy\\
\and
Dipartimento di Fisica, Universit\`{a} degli Studi di Milano, Via Celoria, 16, Milano, Italy\\
\and
Dipartimento di Fisica, Universit\`{a} degli Studi di Trieste, via A. Valerio 2, Trieste, Italy\\
\and
Dipartimento di Fisica, Universit\`{a} di Roma Tor Vergata, Via della Ricerca Scientifica, 1, Roma, Italy\\
\and
Discovery Center, Niels Bohr Institute, Blegdamsvej 17, Copenhagen, Denmark\\
\and
Dpto. Astrof\'{i}sica, Universidad de La Laguna (ULL), E-38206 La Laguna, Tenerife, Spain\\
\and
European Southern Observatory, ESO Vitacura, Alonso de Cordova 3107, Vitacura, Casilla 19001, Santiago, Chile\\
\and
European Space Agency, ESAC, Planck Science Office, Camino bajo del Castillo, s/n, Urbanizaci\'{o}n Villafranca del Castillo, Villanueva de la Ca\~{n}ada, Madrid, Spain\\
\and
European Space Agency, ESTEC, Keplerlaan 1, 2201 AZ Noordwijk, The Netherlands\\
\and
GEPI, Observatoire de Paris, Section de Meudon, 5 Place J. Janssen, 92195 Meudon Cedex, France\\
\and
Helsinki Institute of Physics, Gustaf H\"{a}llstr\"{o}min katu 2, University of Helsinki, Helsinki, Finland\\
\and
INAF - Osservatorio Astronomico di Padova, Vicolo dell'Osservatorio 5, Padova, Italy\\
\and
INAF - Osservatorio Astronomico di Roma, via di Frascati 33, Monte Porzio Catone, Italy\\
\and
INAF - Osservatorio Astronomico di Trieste, Via G.B. Tiepolo 11, Trieste, Italy\\
\and
INAF Istituto di Radioastronomia, Via P. Gobetti 101, 40129 Bologna, Italy\\
\and
INAF/IASF Bologna, Via Gobetti 101, Bologna, Italy\\
\and
INAF/IASF Milano, Via E. Bassini 15, Milano, Italy\\
\and
INFN, Sezione di Roma 1, Universit`{a} di Roma Sapienza, Piazzale Aldo Moro 2, 00185, Roma, Italy\\
\and
IUCAA, Post Bag 4, Ganeshkhind, Pune University Campus, Pune 411 007, India\\
\and
Imperial College London, Astrophysics group, Blackett Laboratory, Prince Consort Road, London, SW7 2AZ, U.K.\\
\and
Infrared Processing and Analysis Center, California Institute of Technology, Pasadena, CA 91125, U.S.A.\\
\and
Institut Universitaire de France, 103, bd Saint-Michel, 75005, Paris, France\\
\and
Institut d'Astrophysique Spatiale, CNRS (UMR8617) Universit\'{e} Paris-Sud 11, B\^{a}timent 121, Orsay, France\\
\and
Institut d'Astrophysique de Paris, CNRS (UMR7095), 98 bis Boulevard Arago, F-75014, Paris, France\\
\and
Institute for Space Sciences, Bucharest-Magurale, Romania\\
\and
Institute of Astro and Particle Physics, Technikerstrasse 25/8, University of Innsbruck, A-6020, Innsbruck, Austria\\
\and
Institute of Astronomy and Astrophysics, Academia Sinica, Taipei, Taiwan\\
\and
Institute of Astronomy, University of Cambridge, Madingley Road, Cambridge CB3 0HA, U.K.\\
\and
Institute of Theoretical Astrophysics, University of Oslo, Blindern, Oslo, Norway\\
\and
Instituto de Astrof\'{\i}sica de Canarias, C/V\'{\i}a L\'{a}ctea s/n, La Laguna, Tenerife, Spain\\
\and
Instituto de F\'{\i}sica de Cantabria (CSIC-Universidad de Cantabria), Avda. de los Castros s/n, Santander, Spain\\
\and
Jet Propulsion Laboratory, California Institute of Technology, 4800 Oak Grove Drive, Pasadena, California, U.S.A.\\
\and
Jodrell Bank Centre for Astrophysics, Alan Turing Building, School of Physics and Astronomy, The University of Manchester, Oxford Road, Manchester, M13 9PL, U.K.\\
\and
Kavli Institute for Cosmology Cambridge, Madingley Road, Cambridge, CB3 0HA, U.K.\\
\and
LAL, Universit\'{e} Paris-Sud, CNRS/IN2P3, Orsay, France\\
\and
LERMA, CNRS, Observatoire de Paris, 61 Avenue de l'Observatoire, Paris, France\\
\and
Laboratoire AIM, IRFU/Service d'Astrophysique - CEA/DSM - CNRS - Universit\'{e} Paris Diderot, B\^{a}t. 709, CEA-Saclay, F-91191 Gif-sur-Yvette Cedex, France\\
\and
Laboratoire de Physique Subatomique et de Cosmologie, Universit\'{e} Joseph Fourier Grenoble I, CNRS/IN2P3, Institut National Polytechnique de Grenoble, 53 rue des Martyrs, 38026 Grenoble cedex, France\\
\and
Laboratoire de Physique Th\'{e}orique, Universit\'{e} Paris-Sud 11 \& CNRS, B\^{a}timent 210, 91405 Orsay, France\\
\and
Lawrence Berkeley National Laboratory, Berkeley, California, U.S.A.\\
\and
Max-Planck-Institut f\"{u}r Astrophysik, Karl-Schwarzschild-Str. 1, 85741 Garching, Germany\\
\and
Max-Planck-Institut f\"{u}r Extraterrestrische Physik, Giessenbachstra{\ss}e, 85748 Garching, Germany\\
\and
National University of Ireland, Department of Experimental Physics, Maynooth, Co. Kildare, Ireland\\
\and
Niels Bohr Institute, Blegdamsvej 17, Copenhagen, Denmark\\
\and
Observational Cosmology, Mail Stop 367-17, California Institute of Technology, Pasadena, CA, 91125, U.S.A.\\
\and
Optical Science Laboratory, University College London, Gower Street, London, U.K.\\
\and
SISSA, Astrophysics Sector, via Bonomea 265, 34136, Trieste, Italy\\
\and
School of Physics and Astronomy, Cardiff University, Queens Buildings, The Parade, Cardiff, CF24 3AA, U.K.\\
\and
Space Research Institute (IKI), Profsoyuznaya 84/32, Moscow, Russia\\
\and
Space Research Institute (IKI), Russian Academy of Sciences, Profsoyuznaya Str, 84/32, Moscow, 117997, Russia\\
\and
Space Sciences Laboratory, University of California, Berkeley, California, U.S.A.\\
\and
Special Astrophysical Observatory, Russian Academy of Sciences, Nizhnij Arkhyz, Zelenchukskiy region, Karachai-Cherkessian Republic, 369167, Russia\\
\and
Stanford University, Dept of Physics, Varian Physics Bldg, 382 Via Pueblo Mall, Stanford, California, U.S.A.\\
\and
T\"{U}B\.{I}TAK National Observatory, Akdeniz University Campus, 07058, Antalya, Turkey\\
\and
UPMC Univ Paris 06, UMR7095, 98 bis Boulevard Arago, F-75014, Paris, France\\
\and
Universit\'{e} Denis Diderot (Paris 7), 75205 Paris Cedex 13, France\\
\and
Universit\'{e} de Toulouse, UPS-OMP, IRAP, F-31028 Toulouse cedex 4, France\\
% ...
}

\abstract{ We present the scaling relation between Sunyaev-Zeldovich (SZ) signal
  and stellar mass for almost 260,000 locally brightest galaxies (LBGs) selected
  from the Sloan Digital Sky Survey (SDSS).  These are predominantly the central
  galaxies of their dark matter halos.  We calibrate the stellar-to-halo mass
  conversion using realistic mock catalogues based on the Millennium Simulation.
  Applying a multi-frequency matched filter to the \textit{Planck} data for each
  LBG, and averaging the results in bins of stellar mass, we measure the mean SZ
  signal down to $M_\ast\sim 2\times 10^{11}\, \Msolar$, with a clear indication
  of signal at even lower stellar mass. We derive the scaling relation between
  SZ signal and halo mass by assigning halo properties from our mock catalogues
  to the real LBGs and simulating the \textit{Planck} observation process.  This
  relation shows no evidence for deviation from a power law over a halo mass
  range extending from rich clusters down to $M_{500}\sim 2\times 10^{13}\,
  \Msolar$, and there is a clear indication of signal down to $M_{500}\sim
  4\times 10^{12}\, \Msolar$.  \Planck's SZ detections in such low-mass halos
  imply that about a quarter of all baryons have now been seen in the form of
  hot halo gas, and that this gas must be less concentrated than the dark matter
  in such halos in order to remain consistent with X-ray observations.  At the
  high-mass end, the measured SZ signal is 20\,\% lower than found from
  observations of X-ray clusters, a difference consistent with Malmquist bias
  effects in the X-ray sample.}

\date{Received XXXX; accepted YYYY}


\authorrunning{Planck Collaboration}

\maketitle
\section{Introduction}

Galaxy evolution is currently understood to reflect a thermal cycle operating
between baryonic components confined in dark matter halos.  Gas cools
radiatively during the hierarchical build-up of the halo population and
condenses to form galaxies in halo cores. Left unchecked, cooling results in
more massive galaxies than observed \citep{balogh2001, lin2004, tornatore2003},
and one must invoke an additional source of non-gravitational heating to prevent
a ``cooling crisis'' \citep{whiteandrees1978, cole1991, whiteandfrenk1991,
  blanchard1992}. Feedback from star formation and supernovae appears
insufficient to halt cooling in massive halos \citep{borgani2004}, so some
modelers have invoked additional heating by active galactic nuclei
\citep[AGN,][]{churazov2002, springel2005, mcnamara2007}. Such models show
substantially improved agreement with the luminosity-temperature relation of
X-ray clusters \citep{valageas1999, bower2001,cavaliere2002} and the luminosity
function of galaxies \citep{croton2006, bower2006, somerville2008}.  The
energetics of AGN feedback imply that it should have especially strong effects
on low-mass clusters, heating gas in the central regions and pushing it to
larger radii, thereby reducing both gas fractions and X-ray luminosities
\citep{puchwein2008, mccarthy2010}.

Relationships between the gas, stellar, and dark matter properties of halos are
important to our understanding of galaxy formation.  Measurements of these
relationships over a wide range of halo mass, from rich clusters down to
individual galaxies, are therefore a primary objective of a number of current
observational campaigns.  Recent studies have probed the relationship between
the mass of a halo and the stellar mass of its central galaxy (the SHM relation)
using ``abundance matching'' techniques, the dynamics of satellite galaxy
populations, and gravitational lensing \citep{guo2010, moster2010,
  mandelbaum2006a, leauthaud2011}.

Corresponding constraints on the gas content of halos over a similar mass range
are not yet available.  Although there are many detailed X-ray studies of the
intracluster medium, these mostly concern massive clusters; lower mass groups
are faint and so are difficult to study individually.  The Sunyaev-Zeldovich
(SZ) effect \citep{sz1972, birkinshaw1999} offers a fresh means to address this
problem.  Large-area SZ surveys are just beginning to be amassed by ground-based
instruments such as the Atacama Cosmology Telescope \citep[ACT, ][]{swetz2008,
  marriage2010, sehgal2010, hand2011}, the South Pole Telescope
\citep[SPT,][]{carlstrom2009, staniszewski2009, vanderlinde2010, williamson2011}
and APEX-SZ \citep{dobbs2006}, as well as by the \Planck\footnote{
  \Planck\ (\url{http://www.esa.int/Planck}) is a project of the European Space
  Agency (ESA) with instruments provided by two scientific consortia funded by
  ESA member states (in particular the lead countries France and Italy), with
  contributions from NASA (USA) and telescope reflectors provided by a
  collaboration between ESA and a scientific consortium led and funded by
  Denmark.}  satellite mission, \citep{planck2011-5.1a, planck2011-5.1b,
  planck2011-5.2a, planck2011-5.2b, planck2011-5.2c}.

High S/N observations of individual objects are not currently possible over the
full mass range from galaxy clusters down to individual bright galaxies.  The
SHM relation can only be estimated for lower mass objects through statistical
methods applied to large catalogues.  In this context, the SZ effect presents
exciting new opportunities.  First steps in this direction were taken by
\citet{planck2011-5.2c} and \citet{hand2011}, with more recent work by
\citet{draper2012} and \citet{sehgal2012}.  In our first study
\citep{planck2011-5.2c}, we binned large numbers of maxBCG \citep{koester2007b}
clusters by richness to measure the relation between mean SZ signal and
richness.  In a similar manner, \citet{hand2011} binned ACT measurements of
luminous red galaxies to determine the mean relation between SZ signal and LRG
luminosity.

Here, we extend our previous work with \Planck\ multi-frequency observations of
a large sample of locally brightest galaxies (LBGs).  These were selected from
the Sloan Digital Sky Survey (SDSS) using criteria designed to maximize the
fraction of objects that are the central galaxies of their dark matter halos.
We stack the \Planck\ data in order to estimate the mean SZ signal for LBGs in a
series of stellar mass bins.  We then use mock galaxy catalogues based on the
Millennium Simulation and tuned to fit the observed abundance and clustering of
SDSS galaxies to establish the relation between stellar and halo mass.
\Planck\ is a unique SZ instrument for this purpose because of its large
frequency coverage and the fact that it observes the entire SDSS survey area,
allowing study of large samples of galaxy systems with extensive
multi-wavelength data.

We unambiguously ($>3\sigma$) detect the SZ signal down to stellar masses of
$2\times10^{11}\, \Msolar$, corresponding to an effective halo mass $M_{500}$
of $2\times10^{13}\,\Msolar$ (see Sect.~\ref{sec:data}) and we find clear
indications of signal down to $10^{11}\,\Msolar$ ($M_{500}= 4\times 10^{12}\,
\Msolar)$. Detailed simulation both of the galaxy sample and of the
\Planck\ measurement process allows us to correct the effects of halo
miscentering and of the scatter in halo mass at fixed stellar mass when
estimating the SZ signal-halo mass relation.  We find that the relation is well
described by a single power law within its statistical uncertainties.  At the
high end, our results overlap the mass range probed by X-ray clusters, where we
find a 20\% lower SZ signal than obtained from fits to X-ray selected cluster
samples.  This difference is consistent with possible Malmquist bias effects in
the X-ray sample.  The gas properties of dark matter halos display a remarkable
regularity from the poorest groups to the richest clusters.


Throughout this paper, we adopt a fiducial $\Lambda$CDM cosmology consistent
with the WMAP7 results \citep{komatsu2010}. In particular, we use $\Omega_{\rm
  m} = 0.272$, $\Omega_\Lambda=0.728$, $n_s=0.961$, and $\sigma_8 = 0.807$. We
express the Hubble parameter at redshift $z$ as $H(z) = H_0 E(z)$, with $H_0 = h
\times 100$\,km\,s$^{-1}$\,Mpc$^{-1}$ and $h = 0.704$. For the redshift range of
interest ($z \la 1$), we approximate $E^2(z) = \Omega_{\rm m} (1+z)^3 +
\Omega_{\Lambda}$.  The virial radius of a halo is defined here as $R_{200}$,
%
%
the radius enclosing a mean density 200 times the critical density at that redshift, i.e., 
$200\times \rho_{\rm c}(z)$, where $\rho_{\rm c}(z) = 3H^2(z)/(8\pi G)$.
The virial mass is then defined as 
$$M_{200} \equiv 200 (4\pi/3)R_{200}^3\, \rho_{\rm{c}},$$
which we also refer to as $M_{\rm h}$.
Similarly, we quote the conventional masses $M_{500}$ and radii, $R_{500}$, when
presenting the SZ scalings.  For stellar mass, we use the symbol $M_\ast$.

The SZ signal is characterized by $Y_{500}$, the Comptonization parameter
integrated over a sphere of radius $R_{500}$, expressed in square arcminutes.
Specifically,
$$Y_{500} \equiv (\sigma_{\rm T}/(m_{\rm e}c^2)) \int_{0}^{R_{500}} PdV/D_{\rm A}^2(z),$$
where $D_{\rm A}(z)$ is the angular-diameter distance, $\sigma_{\rm T}$ is the
Thomson cross-section, $c$ is the speed of light, $m_{\rm e}$ is the electron
rest mass, and $P=n_{\rm e} k T_{\rm e}$ is the pressure, obtained as the
product of the electron number density and the electron temperature.  Throughout
this paper, we use the quantity
$$\Yscaled \equiv Y_{500}E^{-2/3}(z)(D_{\rm A}(z)/500~{\rm Mpc})^2,$$ 
also expressed in square arcminutes, as the intrinsic SZ signal, scaled to
redshift $z=0$ and to a fixed angular diameter distance.

The paper is organized as follows. Sect.~\ref{sec:data} describes the
\Planck\ maps used in our analysis, and our reference catalogue of locally
brightest galaxies, based on SDSS data. Sect.~\ref{sec:analysis} describes our
methodology.  Sects.~\ref{sec:results} and \ref{sec:robustness} gives our main
results and the tests made to demonstrate their robustness.
Sections~\ref{sec:discussion} and \ref{sec:conclusions} contain discussion and
conclusions, respectively.

\section{Data}
\label{sec:data}

\subsection{Planck Data Set}

\Planck\ \citep{tauber2010a, Planck2011-1.1} is the third generation space
mission to measure the anisotropy of the cosmic microwave background (CMB).  It
observes the sky in nine frequency bands covering 30--857\,GHz with high
sensitivity and angular resolution from 31\arcm\ to 5\arcm.  The Low Frequency
Instrument \citep[LFI;][]{Mandolesi2010, Bersanelli2010, Planck2011-1.4} covers
the 30, 44, and 70\,GHz bands with amplifiers cooled to 20\,\hbox{K}.  The High
Frequency Instrument (HFI; \citealt{Lamarre2010, Planck2011-1.5}) covers the
100, 143, 217, 353, 545, and 857\,GHz bands with bolometers cooled to
0.1\,\hbox{K}.  Polarisation is measured in all but the highest two bands
\citep{Leahy2010, Rosset2010}.  A combination of radiative cooling and three
mechanical coolers produces the temperatures needed for the detectors and optics
\citep{Planck2011-1.3}.  Two data processing centres (DPCs) check and calibrate
the data and make maps of the sky \citep{Planck2011-1.7, Planck2011-1.6}.
\Planck's sensitivity, angular resolution, and frequency coverage make it a
powerful instrument for Galactic and extragalactic astrophysics as well as for
cosmology.  Early astrophysics results are given in Planck Collaboration
VIII--XXVI 2011, based on data taken between 13~August 2009 and 7~June 2010.
Intermediate astrophysics results are now being presented in a series of papers
based on data taken between 13~August 2009 and 27~November 2010.

\subsection{A Locally Brightest Galaxy Catalogue}
\label{sec:sample}

To select a sample of central galaxies, we first define a parent population with
$r<17.7$ ($r$-band, extinction-corrected, Petrosian magnitude) from the
spectroscopic galaxy catalogue of the New York University Value Added Galaxy
Catalogue\footnote{NYU-VAGC, \url{http://sdss.physics.nyu.edu/vagc/}}. This was
built by \cite{2005AJ....129.2562B} based on the seventh data release of the
Sloan Digital Sky Survey \citep[SDSS/DR7][]{2009ApJS..182..543A}. This parent
catalogue contains 602,251 galaxies.  We then define ``locally brightest
galaxies" to be the set of all galaxies with $z>0.03$ that are brighter in $r$
than all other sample galaxies projected within 1.0\,Mpc
and with redshift differing by less than 1,000\,km\,s$^{-1}$. After this cut
347,486 locally brightest galaxies remain.

The SDSS spectroscopic sample is incomplete because it proved impossible to
place a fibre on every object satisfying the photometric selection criteria, and
because some spectra failed to give acceptable redshifts. The completeness to
our chosen magnitude limit varies with position, with a mean of 91.5\,\% over
the survey as a whole. To ensure that galaxies without SDSS spectroscopy do not
violate our sample selection criteria, we have used SDSS photometry to eliminate
all objects with a companion that is close and bright enough that it might
violate the above criteria. Specifically, we have used the ``photometric
redshift 2'' catalogue \citep[photoz2][]{2009MNRAS.396.2379C} from the SDSS DR7
website to search for additional companions.  This catalogue tabulates a
redshift probability distribution in bins of width $\Delta z = 0.0145$ for every
galaxy down to photometric limits much fainter than we require.  We then
eliminate any candidate with a companion in this catalogue of equal or brighter
$r$-magnitude and projected within 1.0\,Mpc, unless the photometric redshift
distribution of the ``companion'' is inconsistent with the spectroscopic
redshift of the candidate. (Our definition of ``inconsistent'' is that the total
probability for the companion to have a redshift equal to or less than that of
the candidate is less than 0.1; in practice this eliminates ``companions'' that
are too red to be at a redshift as low as that of the candidate.) This procedure
leaves us with a cleaned sample of 259,579 locally brightest galaxies.

\begin{figure*}
\centering
\includegraphics[width=9cm]{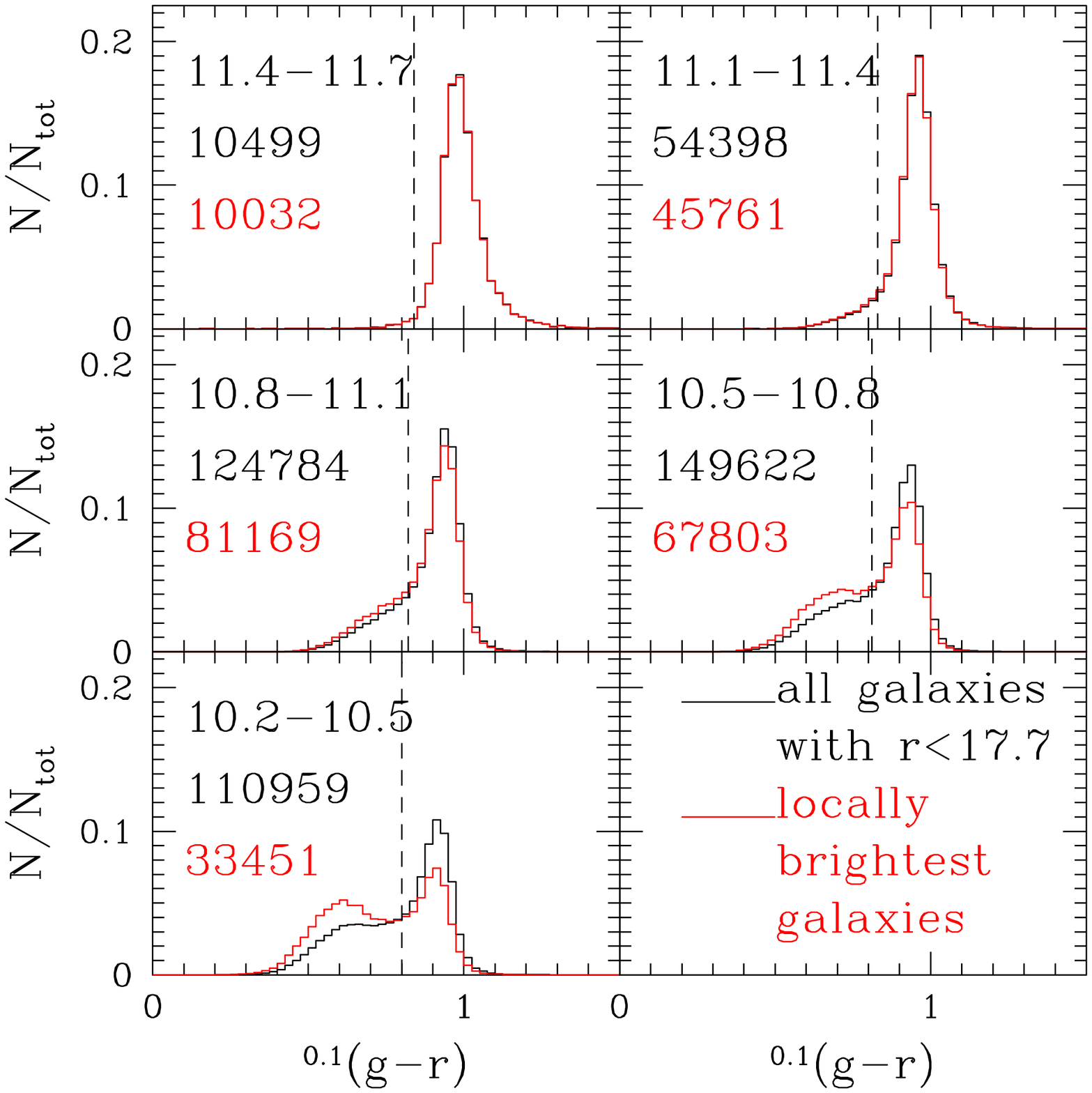}
\includegraphics[width=9cm]{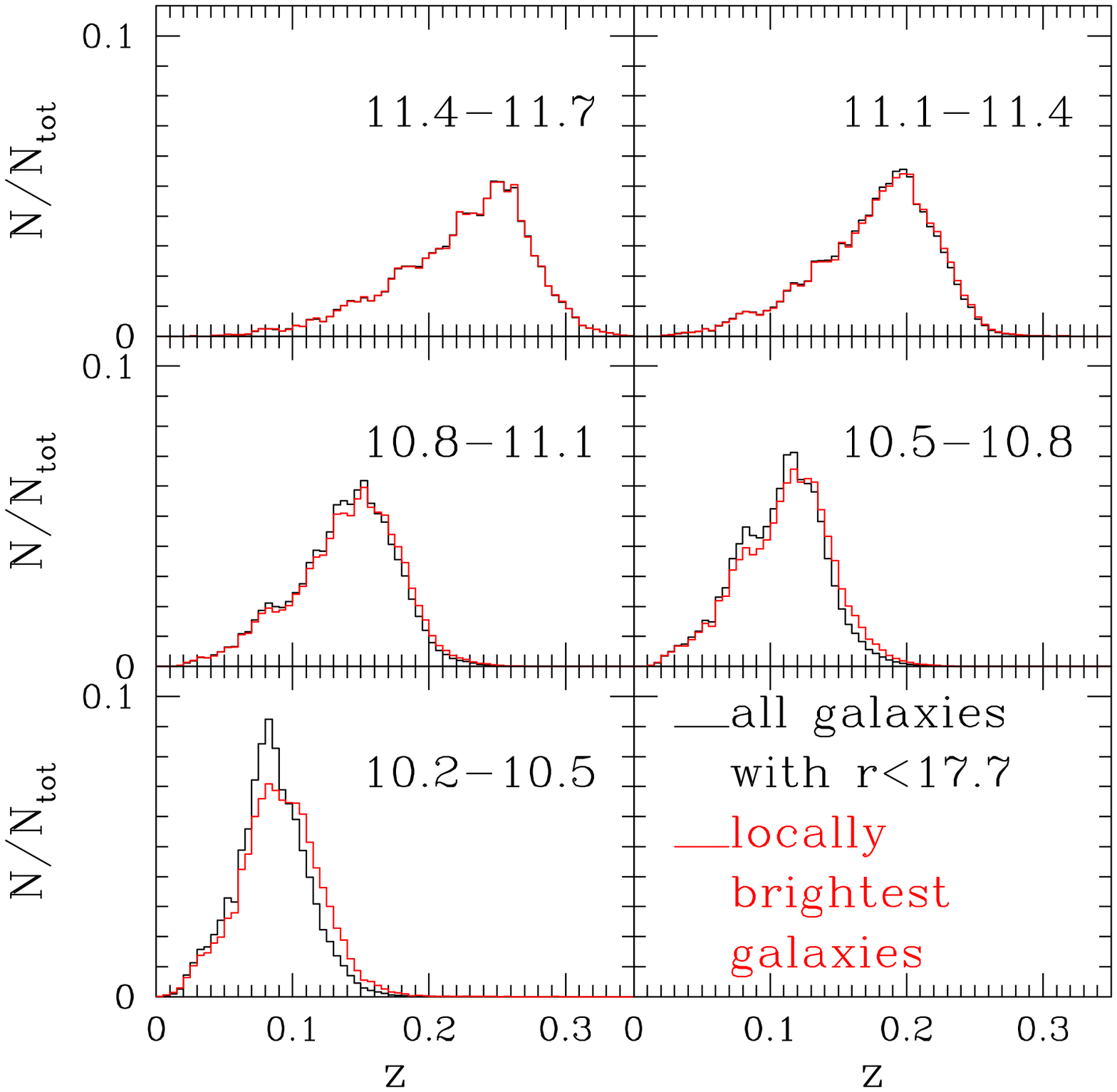}
\caption{Distributions in colour (\emph{left}) and redshift (\emph{right}) of our locally
  brightest galaxies and of the SDSS/DR7 population from which they were
  drawn. Black histograms refer to the parent sample and red histograms to the
  locally brightest galaxies.  The panels in each set correspond to five
  disjoint ranges of $\log_{10} M_\ast/\rm{\Msolar}$, as indicated in the labels. In the
  left-hand set, additional labels give the number of galaxies contributing to
  the parent (black) and locally brightest (red) histograms. Dashed vertical
  lines in these same panels indicate the colour we use to separate red and blue
  galaxies in Fig.~\ref{fig:scatter} below.}
\label{fig:distributions}
\end{figure*}

The NYU-VAGC provides a variety of data for each galaxy. In addition to the
positions, magnitudes, and redshifts used to create our sample, we will make use
of rest-frame colours and stellar masses.  The latter are based on stellar
population fits to the five-band SDSS photometry and on the measured redshifts,
assuming a \cite{2003PASP..115..763C} stellar initial mass function
\citep{2007AJ....133..734B}. In Fig.~\ref{fig:distributions} we compare the
colour and redshift distributions of our final sample of locally brightest
galaxies to those of the parent sample for five disjoint ranges of stellar
mass. For $\log_{10} M_\ast/\rm{\Msolar}\geq 10.8$, the distributions are
similar for the two populations. At lower stellar mass, locally brightest
galaxies are a small fraction of the parent sample and are biased to bluer
colours and to slightly larger redshifts.  In our stacking analysis below, we
obtain significant SZ signals only for galaxies with $\log_{10}
M_\ast/\rm{\Msolar}\geq 11.0$. Our sample contains 81,392 galaxies satisfying
this bound, the great majority of them on or near the red sequence.

\subsubsection{The reliability of our central galaxy sample and its stellar
mass-halo mass relation}
\label{sec:cgc}

We expect the majority of our locally brightest galaxies to be the central
galaxies of their dark matter halos, just as bright field galaxies lie at the
centres of their satellite systems and cD galaxies lie near the centres of their
clusters and are normally their brightest galaxies.  For our later analysis, it
is important to know both the reliability of our galaxy sample, i.e., the
fraction of galaxies that are indeed the central galaxies of their halos, and
the relation between the observable stellar masses of the galaxies and the
unobservable masses of their halos. In this section we investigate both issues
using an update of the publicly
available\footnote{\url{http://www.mpa-garching.mpg.de/millennium}}
semi-analytic galaxy formation simulation of \citet{2011MNRAS.413..101G}. The
update uses the technique of \citet{2010MNRAS.405..143A} to rescale the
Millennium Simulation \citep{2005Natur.435..629S} to the WMAP7 cosmology, then
readjusts the galaxy formation parameters to produce a $z = 0$ galaxy population
with abundance and clustering properties that are almost indistinguishable from
those of the original model. At the relatively high masses relevant for our
work, this simulation provides a very close match to the observed luminosity and
stellar mass functions of the SDSS as well as to the auto-correlations of SDSS
galaxies as a function of stellar mass \citep{guo2012}.

We construct a sample of locally brightest galaxies from this simulation using
criteria exactly analogous to those used for the measured data.  We project the
galaxy distribution onto one of the faces of the simulation cube and assign each
galaxy a redshift based on its distance and peculiar velocity in the projection
direction.  A galaxy is considered locally brightest if it has no neighbour that
is brighter in $r$ within 1.0\,Mpc projected distance and 1,000\,km\,s\mo\ in
redshift.  We divide galaxies into ``centrals'', defined as those lying at the
minimum of the gravitational potential of the dark matter friends-of-friends
(FoF) group with which they are associated, and ``satellites'', defined as all
other galaxies.

\begin{figure}
\centering
\includegraphics[width=9cm]{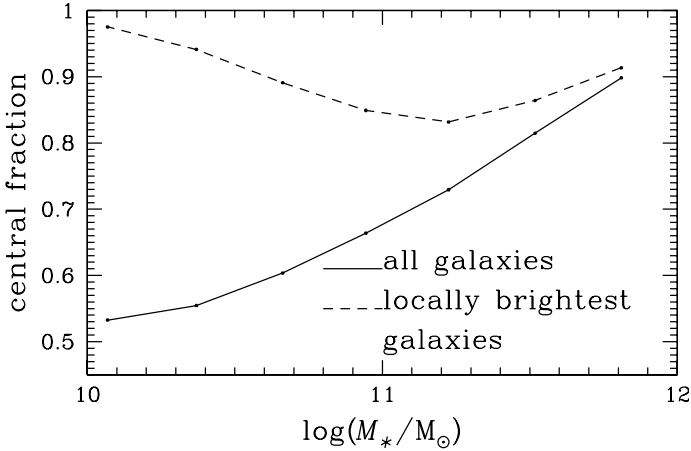}
\caption{Fraction of locally brightest galaxies that are the central objects in
  their dark halos, based on the simulations of \citet{2011MNRAS.413..101G}.
  The solid line traces the fraction of {\it all\/} simulated galaxies that are
  central galaxies as a function of stellar mass.  This fraction increases with
  stellar mass, reaching 90\,\% at the high mass end.  The dashed line presents
  the central galaxy fraction for locally brightest galaxies selected from the
  simulation according to the criteria applied to the SDSS data. This yields a
  sample that is over 83\,\% reliable at all stellar masses. }
\label{fig:cenfrac}
\end{figure}

With these definitions we can assess the fraction of our locally brightest
galaxies that are truly central galaxies. The black line in
Fig.~\ref{fig:cenfrac} shows, as a function of stellar mass, the fraction of
{\it all} galaxies in the simulation that are centrals. At stellar masses just
above $10^{10}\,\Msolar$ this fraction is about one half, but it increases with
stellar mass, reaching two thirds by $\log_{10} M_\ast/\rm{\Msolar} = 11.0$ and
90\,\% by $\log_{10} M_\ast/\rm{\Msolar} =11.8$.  In contrast, the fraction of
locally brightest galaxies that are centrals is much higher, with a minimum of
just over 83\,\% at stellar masses somewhat above $10^{11}\,\Msolar$.  We have
checked those locally brightest galaxies that are satellites, finding that for
$\log_{10} M_\ast/\Msolar>11$, about two-thirds are brighter than the true
central galaxies of their halos. The remainder are fainter than their centrals,
and are considered locally brightest because they are more than 1\,Mpc
(projected) from their centrals (60\,\%) or have redshifts differing by more
than 1,000\,km\,s$^{-1}$ (40\,\%).

\begin{figure}
\centering
\includegraphics[width=9cm]{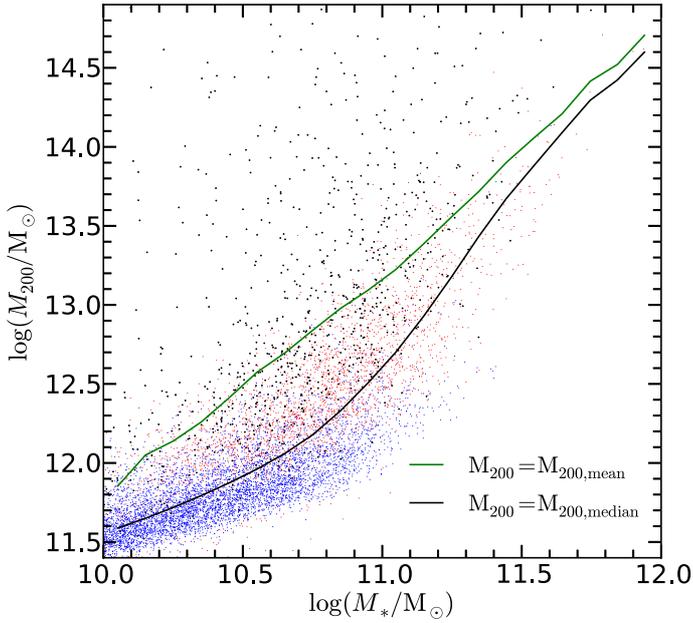}
\caption{Scatter plot of $M_{\rm 200}$ against $M_\ast$ for a random subset (one
  out of 80) of our sample of simulated locally brightest galaxies. Central
  galaxies are shown as red or blue points according to their $g-r$ colour,
  using the cuts indicated in Fig.~\ref{fig:distributions}.  Satellite galaxies
  are shown as black points. The lower and upper curves give the median and mean
  values of halo mass as a function of stellar mass. }
\label{fig:scatter}
\end{figure}

We can assign a halo mass, $M_{\rm 200}$, to every galaxy in our simulation.
For both satellite galaxies and central galaxies, we take $M_{\rm 200}$ to be
the current $M_{200}$ of the FoF dark matter halo with which the object is
associated, i.e., the mass contained within its $R_{200}$.
Figure~\ref{fig:scatter} shows a scatter plot of $M_{\rm 200}$ against $M_\ast$
for a random subset (one out of every 80) of our sample of simulated locally
brightest galaxies.  We indicate central galaxies with red or blue points
according to their rest-frame $g-r$ colour (with the two distinct regions
separated by the vertical dashed lines in the left panel of
Fig.~\ref{fig:distributions}) while satellite galaxies are indicated by black
points. Clearly, red (passive) and blue (star-forming) central galaxies lie on
different $M_{\rm 200}$-$M_\ast$ relations.  That for passive galaxies is
steeper, and is offset to larger halo mass in the stellar mass range where both
types of central galaxy are present.

Satellite galaxies lie in halos in the massive tail of the distribution for
central galaxies of the same stellar mass.  Satellites misidentified as centrals
in our catalogue are usually outlying members projected at relatively large
separation (from a few hundred kiloparsecs to 2\,Mpc).  Their presence bias high
both the mean halo mass (the high black points in Fig.~\ref{fig:scatter}) and
the spatial extent of the stacked SZ signals we measure below.  However, since
two thirds of the satellites that we misidentify as central galaxies are in fact
brighter than the true central galaxies of their halos (i.e., they are not
typical satellites), this bias is not extreme.  In any case, we correct for
these effects explicitly in our analysis using the simulation.

The lower of the two continuous curves in Fig.~\ref{fig:scatter} shows the
median $M_{\rm 200}$ as a function of $M_\ast$. We will take this as an estimate
of the typical halo mass associated with a central galaxy of known $M_\ast$, and
will use it to set the angular size of the matched filter for each observed
galaxy when stacking SZ signal as a function of stellar mass. The upper
continuous curve shows the mean $M_{\rm 200}$ as a function of $M_\ast$.  The
substantial shift between the two is a measure of the skewness induced by the
differing relations for passive and star-forming centrals and by the presence of
the tail of cluster satellite galaxies (see Appendix~\ref{ap:shm}).

\section{Analysis}
\label{sec:analysis}

Our analysis closely follows that presented in \cite{planck2011-5.2a},
\cite{planck2011-5.2b}, and \cite{planck2011-5.2c}, employing as primary method
a multi-frequency matched filter (hereafter MMF) optimized in both frequency and
angular space to extract the thermal SZ signal \citep{herranz2002,melin2006}.
We find that dust emission from our target sources affects the MMF measurements
noticeably at low stellar mass, and that an effective mitigation is to restrict
our final measurements to the three lowest HFI frequencies (100, 143, and
217~GHz).  This is detailed in Sect.~\ref{sec:robustness}.  Our primary
scientific results are hence all based on this three-band MMF.

For the SZ model template, we employ, as in earlier work
\citep{planck2011-5.2a,planck2011-5.2b,planck2011-5.2c}, the so-called
``universal pressure profile'' \citep{arnaud2010} deduced from X-ray
observations of the REXCESS cluster sample \citep{bohringer2007}. The $R_{500}$
value associated with the halo of each central galaxy is obtained as follows.
We first use the SHM relation giving the median halo $M_{200}$ as a function of
central galaxy stellar mass, as presented in Sect.~\ref{sec:cgc}. Then, using an
NFW profile \citep{NFW} and the concentration parameter $c_{200}$ given by
\cite{neto}, we convert $M_{200}$ to $M_{500}$ and derive $R_{500}$ for each
halo.  The angular scale for the filter is finally given by projecting $R_{500}$
at the redshift of the target LBG.

In addition to the MMF, and in order to test the robustness of the results, the
impact of foreground contamination and possible systematic effects, we have also
implemented aperture photometry (hereafter AP).  For the AP, given an object of
certain angular size $R$, the method evaluates the mean temperature in a circle
of radius $r=R$ and subtracts from it the average found in a surrounding ring of
inner and outer radii $r=R$ and $r = f R$, respectively, with $f>1$ \citep[see
  e.g.,][]{2004MNRAS.347..403H}. By removing the mean temperature in the outer
region, the method corrects for large-scale fluctuations in the background. Once
the temperature estimates are derived for each frequency map, they are combined
with inverse-variance weighting to derive an SZ signal estimate by using the
known frequency dependence of the (non-relativistic) thermal SZ effect. Our
choice for the two parameters of the AP method is $(R,f) =
(\rm{FWHM},\sqrt{2})$. Note that the FWHM varies from one frequency band to
another. We also note that the flux estimates within the aperture have to be
corrected separately at each frequency by an appropriate factor in order to
obtain the total flux of the source. For example, if the objects are unresolved
and we assume Gaussian beams, then this correction factor is $(1+ \exp(-8\ln 2)
- \exp(-4 \ln 2))^{-1}$ for the above choice of $R$ and $f$. For extended
objects (e.g., those objects with $R_{500}$ larger than the beam size, and which
are modeled here using the ``universal pressure profile''), the conversion
factor can be evaluated numerically.

Using one of these methods (MMF or AP), we obtain a measure of the intrinsic SZ
signal strength $\Yscaled(i)$ and the associated measurement uncertainty
$\sigscaled(i)$ for the halo of each galaxy $i$.  The majority of these
individual SZ measurements have low signal-to-noise ratio.  Following the
approach in \citet{planck2011-5.2a} and \citet{planck2011-5.2c}, we bin them by
stellar mass, calculating the bin-average signal $\langle \Yscaled
\rangle_{b}=[\sum
  1/\sigscaled^2(i)]^{-1}\sum_{i=1}^{N_{b}}\Yscaled(i)/\sigscaled^2(i)$, with
uncertainty $\sigma^{-2}_{b}=\sum_{i=1}^{N_{b}}1/\sigscaled^2(i)$, where $N_{b}$
is the number of galaxies in bin $b$.

%

\section{Results}
\label{sec:results}

\begin{figure}
\centering
\includegraphics[width=9cm]{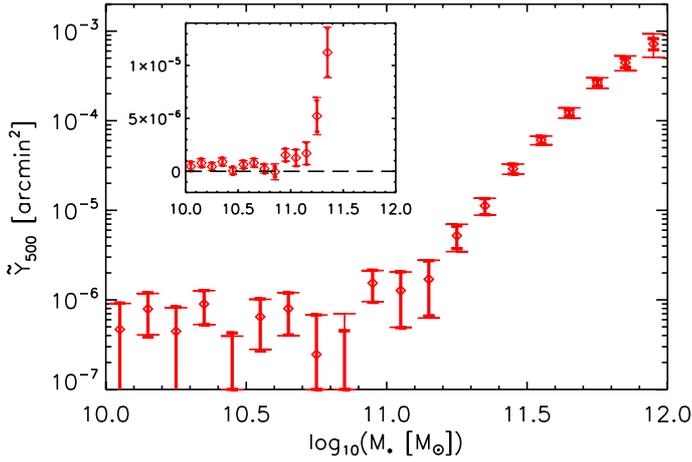}
\caption{Mean SZ signal vs.~stellar mass for locally brightest galaxies.  
  Thick error bars trace the uncertainty on the bin
  average due purely to measurement error, while thin bars with large
  terminators show the variance calculated by bootstrap resampling and so also
  include the intrinsic scatter in the signal.  The inset provides a view on a
  linear scale to better evaluate the significance of the detections.  We
  observe a clear relation between the mean SZ signal and stellar mass down to
  $\log_{10} (M_\ast/\Msolar) = 11.25$ (the detection in this bin is at
  $3.5\,\sigma$), with a suggestion of signal to lower mass: the next three bins
  show signal at $1.6\,\sigma$, $1.6\,\sigma$ and $2.6\,\sigma$, respectively.}
\label{fig:Ybinned}
\end{figure}

\begin{table}[tmb] 
\begingroup 
\newdimen\tblskip \tblskip=5pt
\caption{\Planck\ SZ signal measurements $\Yscaled$ binned by stellar
  mass (adopting a WMAP7 cosmology).  These data are
  displayed in Fig.~\ref{fig:Ybinned}.
}
\label{tab:Ymstar}
\vskip -6mm
\footnotesize 
\setbox\tablebox=\vbox{ %
\newdimen\digitwidth 
\setbox0=\hbox{\rm 0}
\digitwidth=\wd0
\catcode`*=\active
\def*{\kern\digitwidth}
\newdimen\signwidth
\setbox0=\hbox{+}
\signwidth=\wd0
\catcode`!=\active
\def!{\kern\signwidth}
\halign{\hbox to 2.2cm{#\leaderfil}\tabskip=1.0em& 
         \hfil#\hfil&
         \hfil#\hfil\tabskip=1.0em& 
         \hfil#\hfil\tabskip=0pt\cr
\noalign{\doubleline}
\omit&&\multispan2\hfil E{\sc rrors} [$10^{-6}$\,arcm$^2$]\hfil\cr
\noalign{\vskip -3pt}
\omit& $\Yscaled$&\multispan2\hrulefill\cr
\omit\hfil $\log_{10}\left({M_\ast\over\rm{\Msolar}}\right)$\hfil&[$10^{-6}$\,arcm$^2$]\hfil& Statistical& Bootstrap\cr
\noalign{\vskip 3pt\hrule\vskip 5pt}
10.05& **0.47&          **$\pm 0.45$&            **$\pm 0.44$\cr
10.15& **0.79&          **$\pm 0.41$&            **$\pm 0.39$\cr
10.25& **0.44&          **$\pm 0.39$&            **$\pm 0.37$\cr
10.35& **0.90&          **$\pm 0.37$&            **$\pm 0.37$\cr
10.45& **0.05&          **$\pm 0.37$&            **$\pm 0.34$\cr
10.55& **0.65&          **$\pm 0.38$&            **$\pm 0.37$\cr
10.65& **0.80&          **$\pm 0.39$&            **$\pm 0.40$\cr
10.75& **0.25&          **$\pm 0.43$&            **$\pm 0.43$\cr
10.85&**\llap{$-$}0.05& **$\pm 0.50$&            **$\pm 0.75$\cr
10.95& **1.54&          **$\pm 0.60$&            **$\pm 0.58$\cr
11.05& **1.27&          **$\pm 0.78$&            **$\pm 0.78$\cr
11.15& **1.7*&          **$\pm 1.0$*&            **$\pm 1.1$*\cr
11.25& **5.2*&          **$\pm 1.5$*&            **$\pm 1.8$*\cr
11.35& *11.2*&          **$\pm 2.3$*&            **$\pm 2.4$*\cr
11.45& *29.0*&          **$\pm 3.6$*&            **$\pm 3.8$*\cr
11.55& *60.7*&          **$\pm 6.2$*&            **$\pm 6.8$*\cr
11.65& 123\phantom{.}**& *$\pm 11$\phantom{.}**& *$\pm 16$\phantom{.}**\cr
11.75& 266\phantom{.}**& *$\pm 23$\phantom{.}**& *$\pm 36$\phantom{.}**\cr
11.85& 445\phantom{.}**& *$\pm 53$\phantom{.}**& *$\pm 84$\phantom{.}**\cr
11.95& 721\phantom{.}**& $\pm 103$\phantom{.}**& $\pm 210$\phantom{.}**\cr
\noalign{\vskip 5pt\hrule\vskip 3pt}}}
\endPlancktable 
\endgroup
\end{table}

Our main observational result is given in Fig.~\ref{fig:Ybinned} and
Table~\ref{tab:Ymstar}, showing the mean SZ signal measured using the three-band
MMF for locally brightest galaxies binned according to stellar mass.  In the
plot, the thick error bars show the uncertainty propagated from the individual
measurement errors as described above, while the thin bars with large
terminators give the variance of the weighted bin-average signal found by a
bootstrap resampling.  For the latter, we constructed 1,000 bootstrap
realizations of the original LBG catalogue and performed the full analysis on
each.

The inset uses a linear scale to better display the significance of our
detections.  We have a clear signal down to the bin at $11.2<
\log_{10}(M_\ast/\Msolar) <11.3$, centred at $M_\ast=1.8\times 10^{11}\,
\Msolar$.  The next three bins provide evidence that the signal continues to
lower mass with ``detections'' significant at the $1.6\,\sigma$, $1.6\,\sigma$
and $2.6\,\sigma$ levels, from high to low mass, respectively.  The last bin is
centered at $M_\ast=9\times 10^{10}\, \Msolar$, corresponding to a mean halo
mass of $M_{200}\sim 1.4\times 10^{13}\, \Msolar$.  These last three bins,
however, are more seriously affected by dust contamination, as discussed below,
and for this reason may be more uncertain than these statistical measures
suggest.

\section{Systematic errors}
\label{sec:robustness}

In this section, we present a number of tests of the robustness of our principal
result against systematic error. In the following, unless otherwise stated, all
results use data at 100, 143, and 217\,GHz only.

\subsection{Stacking real-space reconstructed SZ maps}

\begin{figure*}
\centering
\includegraphics[width=18cm]{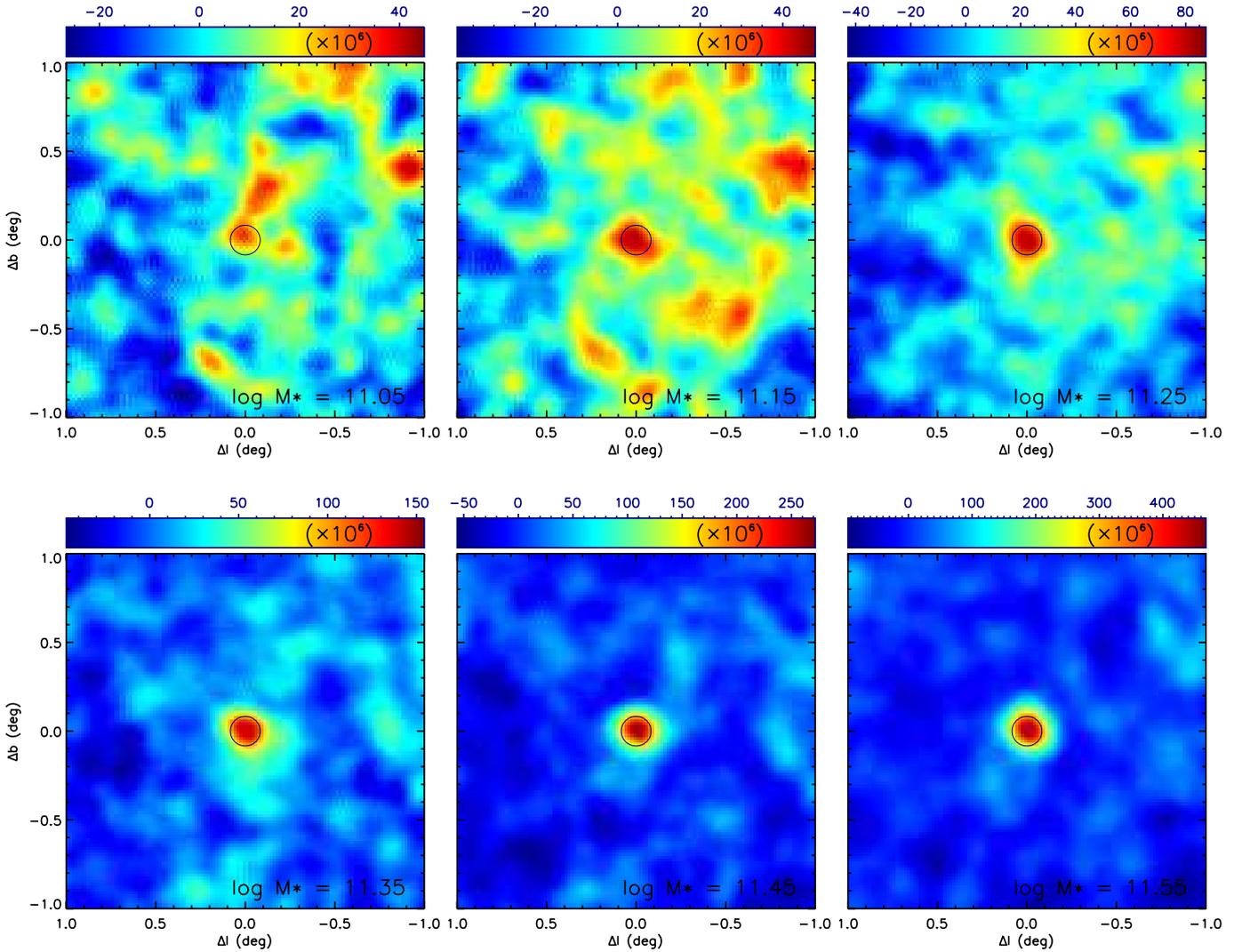}
\caption{Equal-weighted stacks of reconstructed SZ maps (i.e., Comptonization
  parameter maps) for objects in six mass bins centred, from left to right and
  top to bottom, at $\log_{10}(M_\ast/\Msolar) = [11.05, 11.15, 11.25, 11.35,
    11.45, 11.55]$. In all cases, the bin size is taken to be $0.2$, so the
  galaxies in two consecutive panels partially overlap.  Maps are $2\deg$ on a
  side, with Galactic north at the top. The SZ signal traced by the central
  galaxies is clearly detected in all bins above $\log_{10}(M_\ast/\Msolar) =
  11.25$. In all panels, the circles indicate the FWHM of the data, which
  corresponds to $10\arcm$.}
\label{fig:stacking-y-map}
\end{figure*}

According to Fig.~\ref{fig:Ybinned}, the lowest bin at which we have a
$>3\,\sigma$ detection is the one at $\log_{10}(M_\ast/\Msolar) = 11.25$. As a
consistency check, and also as an illustration of the frequency dependence of
the detected SZ signal, Fig.~\ref{fig:stacking-y-map} shows stacked images of
central galaxies in six different mass bins of width $\Delta\log_{10}
M_\ast=0.2$ centred at $\log_{10}(M_\ast/\Msolar) = 11.05, 11.15, 11.25, 11.35,
11.45, \hbox{and } 11.55$.

The stacked maps are obtained, using equal weights, from a (full-sky) SZ map
constructed from the \Planck\ 100, 143, 217, and 353\,GHz maps using a modified
internal linear combination algorithm \citep[MILCA,][]{hurier2010} that has been
used for other \Planck\ Intermediate Papers
\citep[e.g.,][]{planck2012-V,planck2012-X}.  The well-known internal linear
combination approach \citep[e.g.,][]{eriksen04} searches for the linear
combination of the input maps that minimises the variance of the final
reconstructed map while imposing spectral constraints. This preserves the
thermal SZ signal and removes the CMB contamination (using the known spectral
signatures of the two components) in the final SZ map.  The resulting map used
for this analysis has an angular resolution (FWHM) of 10\arcm.  We have checked
that almost identical maps are obtained with other methods.

The SZ signal is clearly visible in all panels with $\log_{10}(M_\ast/\Msolar)
\ge 11.25$. The stacked maps show no sign of a gradient in the residual signal
in the vertical direction, showing that the MILCA method is very effective in
removing Galactic emission.  Below the mass limit of 11.25, there is also some
evidence of SZ signal, although here the contrast relative to the noise is
lower.  Finally, we note that the signal in the lower stellar mass panels is
extended. This is mainly due to the larger satellite fraction at these masses
that results in a significant contribution to the stack from relatively massive
halos with centres significantly offset from the locally brightest galaxy (see
Figs.~\ref{fig:cenfrac}, \ref{fig:scatter} and Appendix~\ref{ap:miscenter}). We
also discuss in Appendix~\ref{ap:miscenter} the impact of dust contamination on
these maps.


\subsection{Null tests}

\begin{figure}
\centering
\includegraphics[width=9cm]{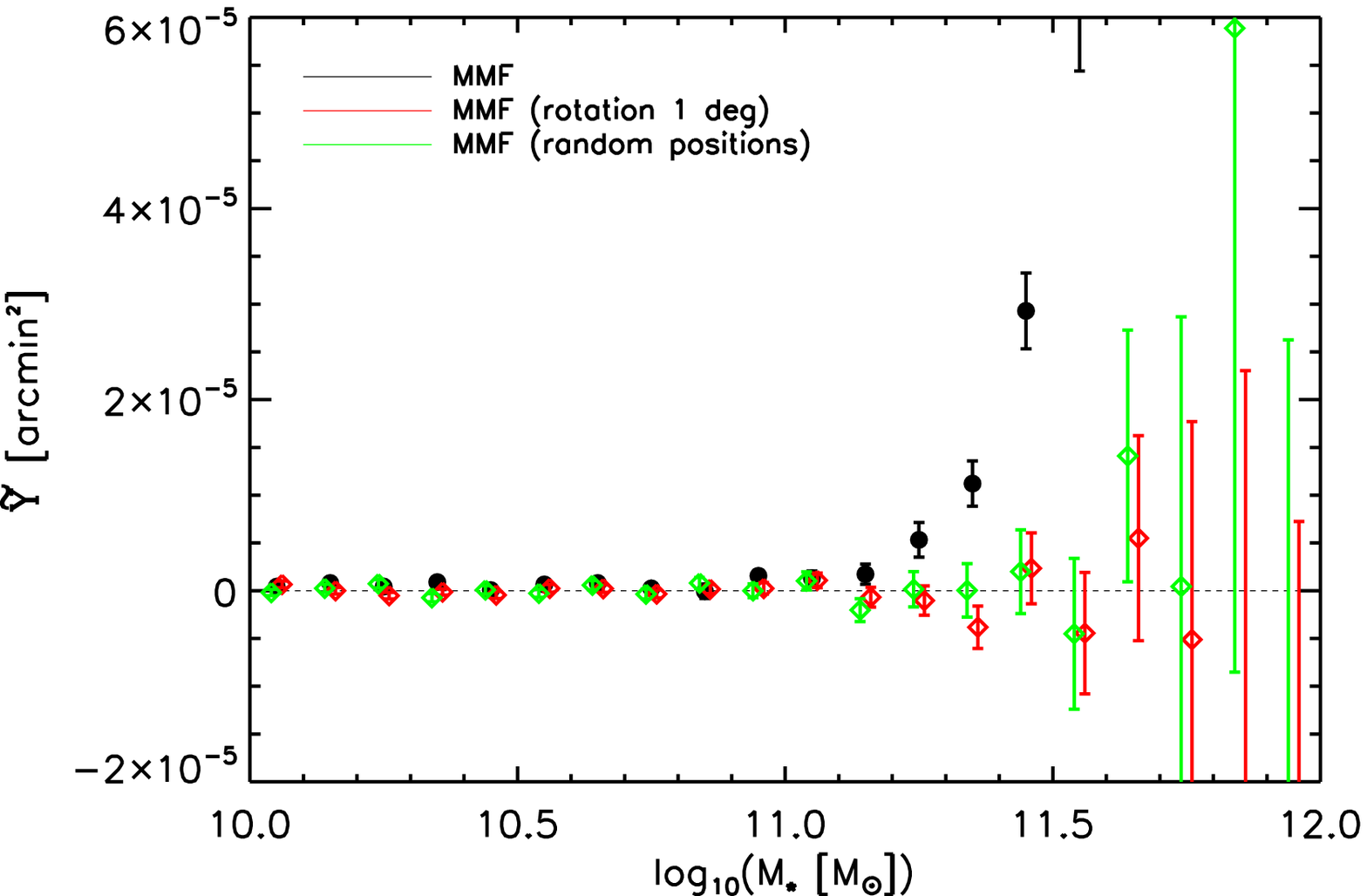}
\caption{Null tests performed on the locally brightest galaxy sample.  Red
  points correspond to placing the filter one degree in declination away from
  the position of each LBG, while green points correspond to random high
  latitude filter positions.  Both sets are consistent with zero.  The black
  points show our measurements with filters centred on the LBG sample,
  demonstrating highly significant detections.  }
\label{fig:null}
\end{figure}

 Null tests used to check for systematic errors are shown in
 Fig.~\ref{fig:null}. Taking the set of MMF filters adapted to each target
 galaxy, we shift their positions on the sky, either with a random displacement
 (i.e., by generating a random distribution of new positions isotropically
 distributed outside our Galactic mask) or by shifting all coordinates one
 degree in declination, and rerunning our analysis.  In both cases the result
 should be zero. The shifted filter sets indeed have bin-average SZ signals
 consistent with zero over the entire mass range.

\subsection{Size effects}

Although most of the halos traced by our locally brightest galaxies are,
according to their inferred $R_{500}$ values, at most marginally resolved by the
\Planck\ beams, size effects are not negligible, and the full pressure profile
has to be used for the flux determination. If instead the objects are
(incorrectly) assumed to be point-like, we find that the flux is underestimated
by roughly 20--30\,\%, although the slope of the $\Yscaled$-$M_\ast$ scaling
relation is practically unaffected.


\subsection{Photometry comparison} 

\begin{figure}
\centering
\includegraphics[width=9cm]{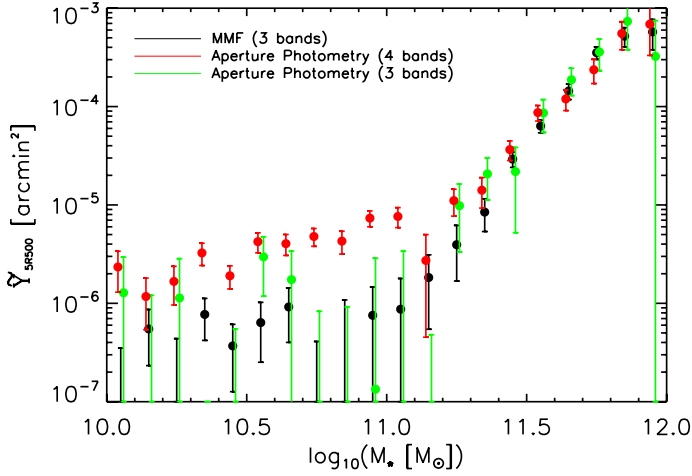}
\caption{Comparison of the SZ measurements on the full locally brightest galaxy
  sample for two different photometry methods: the matched multi-filter (MMF)
  and the aperture photometry (AP) approach. For this figure and only for this
  figure, we assume point-like objects for both methods and plot the derived
  total SZ flux (or the flux within $5R_{500}$ for MMF). The signal detected by
  the two methods is consistent at all stellar masses when only three
  frequencies are used, but when four frequencies are used, the AP results are
  contaminated by high frequency emission at stellar masses below $\sim 2 \times
  10^{11}$\,$\Msolar$. }
\label{fig:photometry}
\end{figure}

Figure~\ref{fig:photometry} compares the SZ signal extracted for our LBG sample
by the two photometry methods described above, namely MMF and \hbox{AP}. Here,
we compare the total SZ flux from MMF (computed as $\tilde{Y}_{5R_{500}}$) with
the total flux recovered from the AP method, after applying the correction
factors described in Sect.~\ref{sec:analysis}. For simplicity, we assume
point-like objects for the flux extraction in this analysis.  This is why the
MMF data points differ from the corresponding points in
Fig.~\ref{fig:Ybinned}. For the AP, we also compare the nominal four-band
analysis with a three-band case to illustrate the impact of residual foregrounds
on our flux estimates.  When the 353\,GHz channel is included, the AP flux
estimates at low stellar-mass are biased towards high SZ values.  This indicates
contaminating high-frequency emission associated with the sources, presumably
dust in the LBGs or their satellites. We discuss this issue further in
Sect.~\ref{sec:dustcontamination}.

The main conclusion is that the two methods, despite their different data
processing approaches, produce fully consistent results for $\log_{10}(M_\ast /
\Msolar) \ga 11.25$, while the results start to show a dependence on the method
for stellar masses below that limit.


\subsection{Dust contamination}
\label{sec:dustcontamination}

\begin{figure}
\centering
\includegraphics[width=9cm]{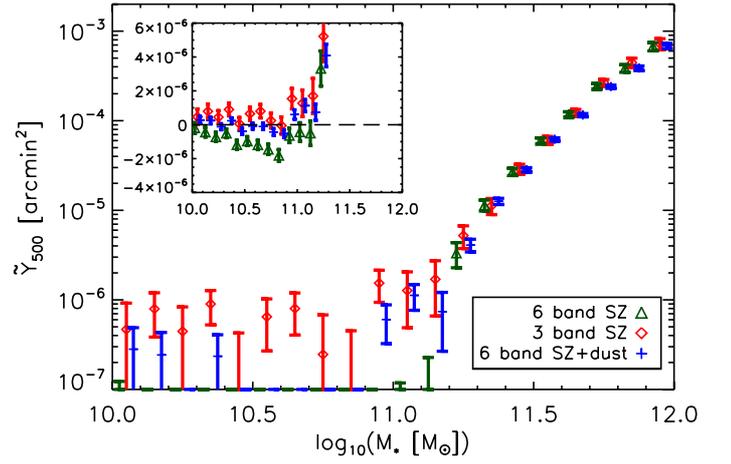}
\caption{Impact of dust contamination on our SZ measurements.  Three cases are
  shown: a 6-band MMF (all \Planck\ HFI frequencies) with no explicit allowance
  for a dust contribution (green triangles), a 3-band MMF also with no explicit
  dust modelling (red diamonds); and a modified 6-band MMF that includes an
  amplitude fit to a fiducial dust spectrum (blue crosses).  The error bars
  include measurement uncertainties only.  For stellar masses where we clearly
  detect the signal (i.e., at $\log_{10} M_\ast/\Msolar > 11.25$), the three
  measurements agree, indicating that dust emission does not significantly
  affect those measurements. At lower masses the 3-band results are consistent
  with the 6-band results when dust is explicitly included in the modelling, but
  not otherwise.}
\label{fig:dust}
\end{figure}

The analysis of the last section suggests that our SZ signal estimates may be
contaminated by residual dust emission that increases with frequency and could
bias our primary results.  To evaluate the potential effects, we have performed
measurements using three different MMFs, as shown in Fig.~\ref{fig:dust}.  The
green triangles and red diamonds represent the results of using all six HFI
channels or only the lowest three (100, 143, 217\,GHz), respectively. In both
cases there is no explicit allowance for a possible dust contribution. The blue
crosses show results for a modified six-band MMF that includes amplitude fits
not only to the SZ spectrum, but also to a fiducial thermal dust spectrum.

The three sets of measurements fully agree for the stellar masses for which we
unambiguously detect the SZ signal, $\log_{10} M_\ast / \Msolar \geq
11.25$. This indicates that dust emission does not significantly affect our
results for these stellar mass bins. At lower mass the three-band results and
the dust-corrected six-band results remain consistent, but the six-band results
without explicit dust correction are systematically different. Dust emission is
clearly sufficient to contaminate our six-band filter estimates of SZ signal if
uncorrected, but it does not appear to be a major problem when only the lower
three frequency bands are used. The residual dust contribution estimated from
the scatter and offset of the red and blue points for $\log_{10} M_\ast /
\Msolar < 11.0$ is below $\sim 10^{-6}$\,arcmin$^2$ and so lies comfortably
below our measured signal.

There is a clear indication of signal in the three bins just below $\log_{10}
M_\ast / \Msolar = 11.25$ both for the three-band MMF and for the dust-corrected
six-band \hbox{MMF}. However, the dust-corrected results appear systematically
lower than the (uncorrected) three-band results by an amount similar to that
seen at lower masses where the SZ signal is undetected. Further, the six-band
MMF measurements without dust correction (the green triangles) differ
substantially for these (and all lower) bins. This suggests that dust emission
affects these stellar mass bins noticeably even for the three-band MMF, so the
corresponding points in Fig.~\ref{fig:Ybinned} may be more uncertain than
indicated by their statistical error bars.  Although formally the dust-corrected
six-band MMF would appear to give our most accurate estimates of stacked SZ
signal, we are uncertain whether the fiducial dust spectrum it assumes is
appropriate for these specific sources.  Therefore we conservatively quote
results based on the three-band MMF, using the dust-corrected six-band results
to give an estimate of remaining dust-related systematics.

Finally, we note that residual dust contamination biases the (uncorrected)
six-band MMF signal estimates for $\log_{10} M_\ast / \Msolar < 11$
(Fig.~\ref{fig:dust}) in the opposite direction to the AP signal estimates (see
Fig.~\ref{fig:photometry}). The agreement of the two methods for $\log_{10}
M_\ast / \Msolar > 11.25 $ is thus a further indication of the robustness of our
primary results.

\subsection{Stability of the signal in different sky surveys}

We have also checked that the SZ signal is stable against splitting the
\Planck\, data into complementary subsets.  For instance, the signal obtained
from the maps of the first 6~months of observing time is fully consistent with
that obtained from maps of the second 6~months and the last 3.5~months (of
course the latter has larger error bars due to its smaller sky coverage).

\section{The $Y_{500}$-$M_{500}$ relation}
\label{sec:discussion}

\begin{figure*} 
\centering
\includegraphics[width=9cm]{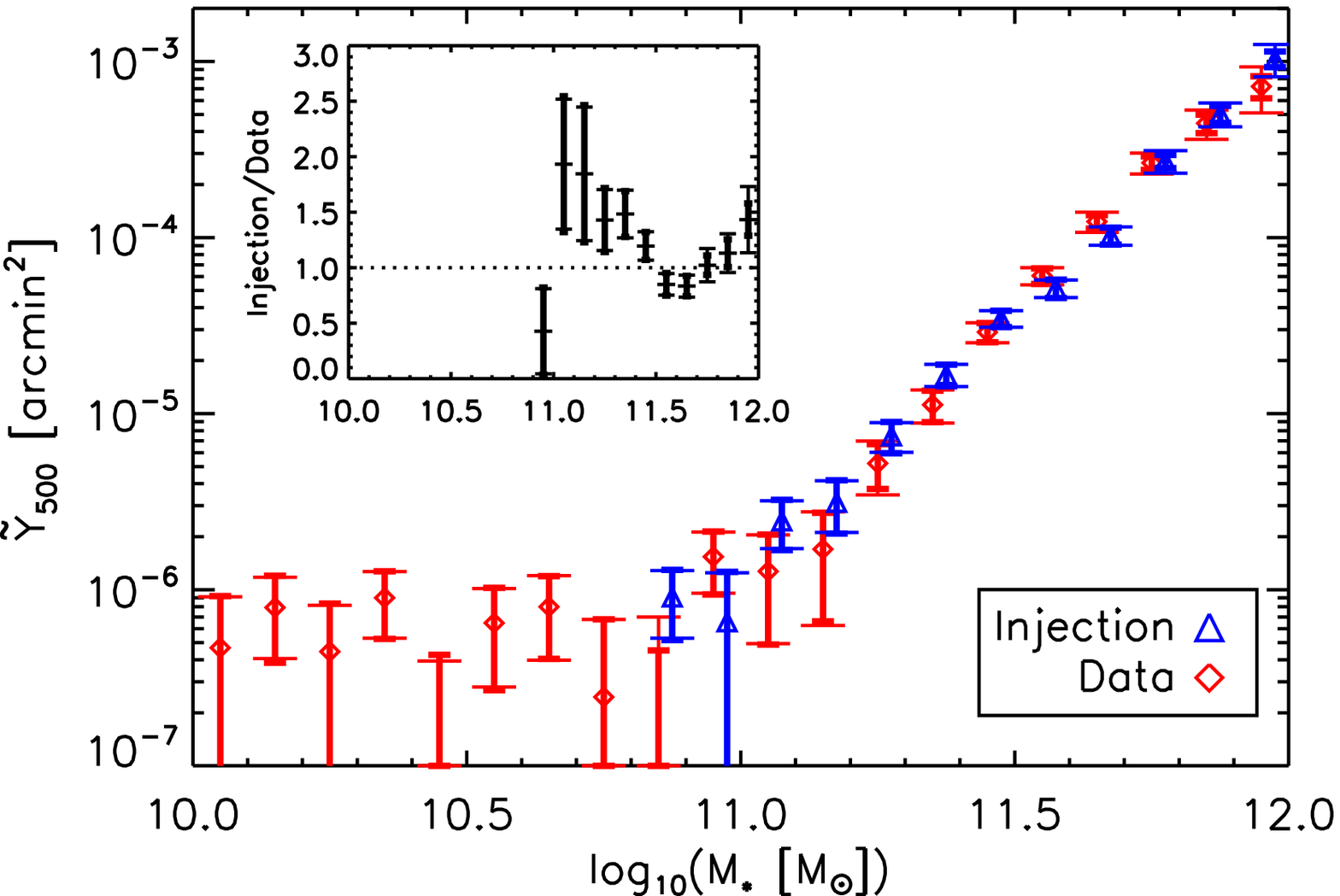}
\includegraphics[trim = 2mm 4mm 12mm 8mm, clip=true,
  width=9cm]{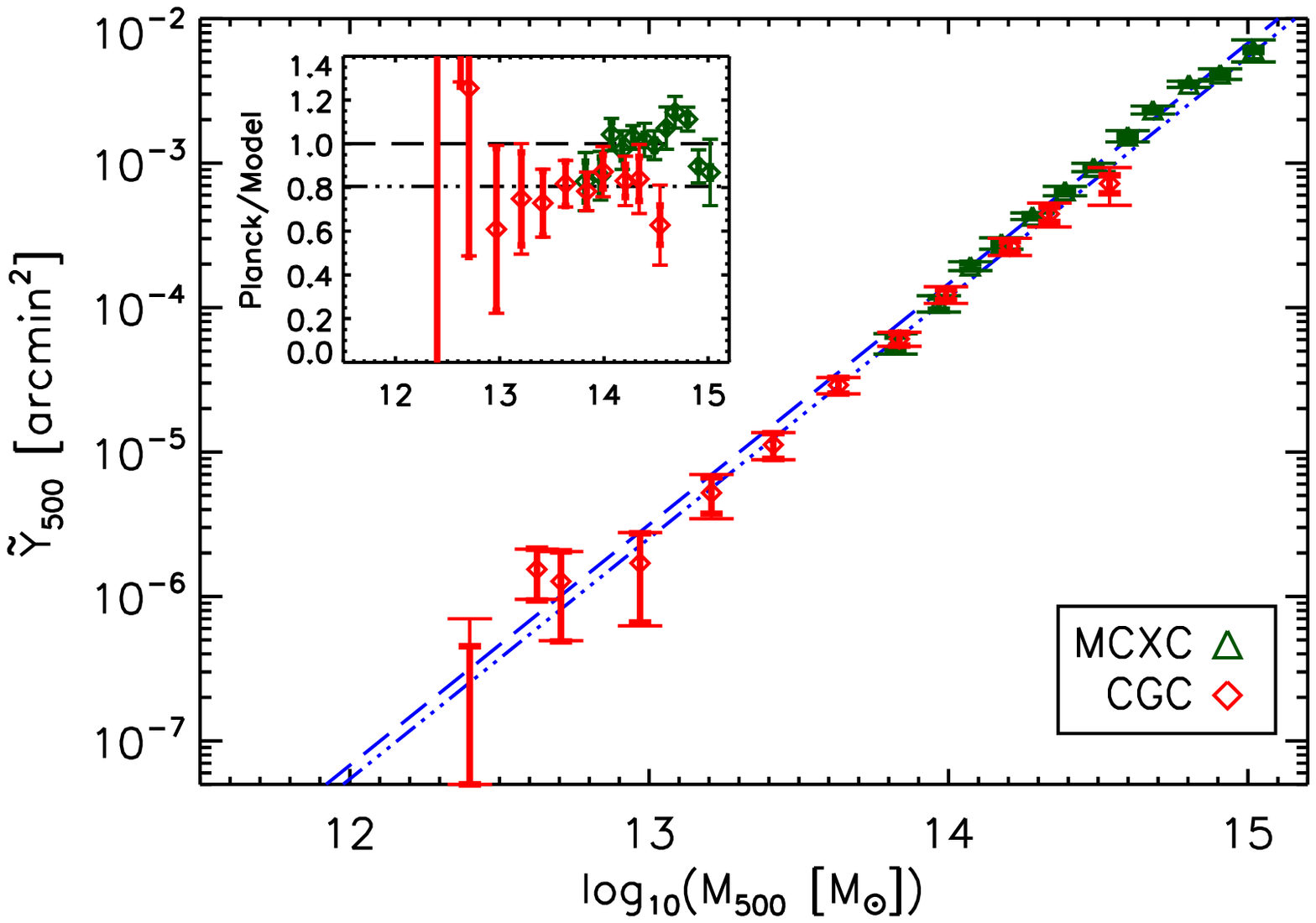}
\caption{{\it Left:} Comparison of the measured mean SZ signal as a function of
  LBG stellar mass (red points) to simulated observations (blue points).  The
  simulations assign to each observed LBG the halo mass and positional offset of
  a randomly chosen simulated LBG of the same stellar mass (compare
  Fig.~\ref{fig:scatter}).  Our best fit $Y_{500}$-$M_{500}$ scaling relation is
  then used, together with the universal pressure profile, to inject a simulated
  signal into the \Planck\ maps (see text). An ``observed'' signal is obtained
  by applying the MMF exactly as for the real data.  The inset gives the ratio
  of the bin-averaged injected and actual signals.  {\it Right:} Mean SZ signal
  as a function of effective halo mass. The bin-averaged SZ signal measurements
  of the left panel have been translated to this plane using the simulations as
  described in the text (the red points). The dot-dashed line is our best fit
  relation between halo mass and SZ signal, i.e., the one leading to the
  simulated measurements in the left panel.  The green points give the mean SZ
  signal of MCXC clusters binned by a halo mass estimated from their X-ray
  luminosity using the REXCESS relation without correction for Malmquist bias
  \citep[line 3 in Table 2 of][]{planck2011-5.2a}.  The dashed blue line shows
  the self-similar model calibrated on the REXCESS sample as given by
  \cite{arnaud2010}.  The inset gives the ratio of all measurements to this
  model's predictions. As in previous figures, the thick error bars account only
  for measurement uncertainties, while thin bars with large terminators result
  from a bootstrap analysis and so include intrinsic scatter effects.}
\label{fig:YM500} 
\end{figure*}

We now turn to the interpretation of our measurements in terms of the SZ
signal-halo mass scaling relation: $Y_{500}$-$M_{500}$.  Our conclusions are
summarised in Fig.~\ref{fig:YM500}.

From our simulation of the locally brightest galaxy catalogue, we expect a large
range of halo masses within a given bin of stellar mass and, in addition, a
fraction of galaxies that are, in fact, satellites, with significant positional
offsets relative to their host halo (see Fig.~\ref{fig:scatter}).  These effects
impact our measurements of the SZ signal-stellar mass relation in two ways.
First, the MMF is not perfectly matched to each individual object because we fix
the filter scale to the median halo size.  This causes an aperture-induced bias
in the flux measurement.  Second, our filter is miscentred for those systems
where the LBG is, in fact, a satellite. These galaxies are often associated with
substantially more massive dark halos than typical LBGs of the same stellar
mass, leading to an increase in the mean signal in the bin, mitigated by the
substantial angular offsets of most such satellites from the true centres of
their rich clusters.  This increases the apparent extent of the signal in
stacked maps like Fig.~\ref{fig:stacking-y-map}, but decreases the contribution
to the signal through a matched filter centred on the galaxy (see Appendix
\ref{ap:miscenter}).

Using our simulation of the LBG catalogue, we can account fully for these
effects and extract the underlying $Y_{500}$-$M_{500}$ relation in an unbiased
way.  Within each stellar mass bin, we identify each observed LBG with a
randomly chosen simulated LBG of the same stellar mass, assigning it the halo
mass and positional offset from halo centre of its partner, but retaining its
observed redshift.  We give each halo a SZ signal distributed according to the
``universal pressure profile'' and normalized using a specific model
$Y_{500}$-$M_{500}$ scaling relation.  Each synthesised object is then observed
with the three-band MMF centred on the galaxy's position, and the measurements
are binned and weighted in the same way as the real data to obtain $\langle Y
\rangle_{\rm s}$.

This procedure enables us to translate a model $Y_{500}$-$M_{500}$ relation to
our observational plane, $Y_{500}$-$M_\ast$, and thus to fit for the underlying
scaling relation with halo mass $M_{500}$.  We model this relation as
\begin{equation}
\label{eq:1}
\Yscaled=\YM\left(\frac{M_{500}}{3\times 10^{14}\, \Msolar} \right)^{\aM},
\end{equation}
fixing the mass exponent to its self-similar value, $\aM = 5/3$, and
fitting for the normalization $\YM$.  Restricting the fit to
$\log_{10}(M_\ast / \Msolar) \ga 11.5$, for direct comparison to X-ray
samples in the discussion below, we find
\begin{equation}
\label{eq:2}
\YM=(0.73\pm 0.07)\times 10^{-3}\, \mbox{\rm arcmin}^2.
\end{equation}   
In the left-hand panel of Fig.~\ref{fig:YM500}, the red points reproduce the
measurements given in Fig.~\ref{fig:Ybinned}, while the blue points show the
simulated observations for this best-fit $Y_{500}$-$M_{500}$ scaling relation.

The best-fit is, however, formally unacceptable, with a reduced $\chi_\nu^2$ of
$3$, which we can more readily appreciate from the inset showing the ratio of
the actual observations to the simulated bin averages on a linear scale.  The
data prefer a shallower slope than the self-similar $\aM=5/3$ over the mass
range of the fit.  Moreover, we see that a power law cannot fit the data over
the full mass range probed by our measurements.  To ease comparison with the
X-ray sample, we will nevertheless adopt this fit below.

The right-hand panel of Fig.~\ref{fig:YM500} considers the SZ signal-halo mass
plane.  The blue dot-dashed line simply traces our best-fit $Y_{500}$-$M_{500}$
relation.  The blue dashed line is the self-similar relation derived from X-ray
cluster studies \citep{arnaud2010}, while the green points present binned SZ
measurements for the approximately 1,600 clusters in the Meta-Catalogue of X-ray
detected Clusters (MCXC) \citep{piffaretti2010}.  The latter measurements are as
reported in \citet{planck2011-5.2a}, with one minor change: in
\citet{planck2011-5.2a} we used an empirical slope for the $Y_{500}$-$M_{500}$
relation taken from X-ray studies; for the points in Fig.~\ref{fig:YM500}, we
repeated the same analysis fixing the slope instead to its self-similar value,
as was done for the LBG sample.  This change moves the green points only very
slightly relative to those shown in \citet{planck2011-5.2a}.  For the mass
estimates of the MCXC objects, we applied the X-ray luminosity-mass relation
from \citet{pratt2009}, corresponding to the case of line 3 of Table 2 in
\citet{planck2011-5.2a}.  The mass is calculated for each MCXC cluster and then
binned.  We plot the point at the median value of the mass in each bin.

To transcribe our central galaxy catalogue measurements onto this figure, we
must first find the effective halo mass corresponding to each stellar mass bin.
This effective mass is a complicated average over the halo masses within the
bin, weighting by the fraction of SZ signal actually observed, i.e., after
accounting for aperture and miscentering effects.  The bin-averaged mean SZ
signals we estimate for our mock LBG catalogue include all these effects, and so
can be used to calculate an effective mass as $M^{\rm eff}_{500} = 3\times
10^{14}\, \Msolar \left(\langle Y \rangle_{\rm s}/\YM\right)^{1/\aM}$, where
$\langle Y\rangle_{\rm s}$ is calculated for each bin as described above, and
$\YM$ and $\aM=5/3$ are the parameters used for Eq.~\ref{eq:1} in the
simulation. (Note that the result is independent of the normalisation $\YM$.)
We do this for a suite of simulated catalogues and take the ensemble average
effective mass for each bin, plotting the results as the red points in the
right-hand panel of the figure.

These LBG results extend the SZ-halo mass scaling relation down in mass by at
least a factor of 3, to $M_{500} = 2\times 10^{13}\,\Msolar$ (the stellar mass
bin at $\log_{10}M_\ast / \Msolar=11.25$).  This is the lowest halo mass for
which the mean SZ signal has been measured.  As previously discussed, there is a
clear indication that the relation continues to even lower mass, with marginally
significant detections in the next three stellar mass bins.  The lowest stellar
mass bin with an apparent SZ detection (at $2.6\sigma$) corresponds to effective
halo mass $\log_{10}M_{500}/\Msolar = 12.6$.  Our power-law fit adequately
describes the data points over more than two orders of magnitude in halo mass
down to this remarkably low value with no hint of a significant deviation.

The inset in the right panel of Fig.~\ref{fig:YM500} shows the ratio of our
measured mean SZ signal to that predicted by the self-similar scaling relation
deduced from X-ray observations of clusters (the dashed blue line
\citep{arnaud2010}). Direct measurements obtained by binning the MCXC clusters
(the green points) agree with this relation. This was the principal result of
\citet{planck2011-5.2a}.  The SZ measurements for our LBGs fall below the
relation, however.  The horizontal dot-dashed line gives the ratio our LBG fit
to the X-ray model (this is the offset between the two blue lines in the main
figure).  Recall that the fit to the LBG catalogue was restricted to masses
overlapping the X-ray sample, $\log_{10}M_{500}/\Msolar >13.8$.  Over this
range, the mean SZ signals associated with LBG halos are about $20\,\%$ lower
than found for X-ray clusters with the same halo mass, a difference that is
significant at the $2.6\,\sigma$ level.

A number of effects could contribute to an offset of this size.  The masses
plotted for the MCXC were calculated using a luminosity-mass relation derived
from the REXCESS sample assuming that halo mass scales self-similarly with the
mass-proxy $Y_X$ and without correction for Malmquist bias \citep{pratt2009}.
Using the Malmquist-corrected relation would remove much of the offset and bring
the two $Y_{500}$-$M_{500}$ scaling relations into acceptable agreement.  In
this sense, the offset is consistent with the estimated effects of Malmquist
bias on the X-ray sample. However, such biases depend on the detailed selection
procedure of the stacked and calibrating cluster samples, on the way in which
the calibration relation is derived, and on the (correlated) intrinsic scatter
of clusters around the $L_x$-$M_{500}$ and $Y_{500}$-$M_{500}$ relations. Thus
they can only be corrected through detailed modelling both of the cluster
population itself and of the definition and analysis of the specific cluster
surveys involved \citep[e.g.,][]{angulo2012}. Furthermore, halo masses are
estimated in very different ways in our two samples --- from X-ray luminosities
calibrated against individual hydrostatic mass measurements for the MCXC, and
through an abundance matching argument based on the WMAP7 cosmology for the LBG
catalogue. Any offset between these two halo mass scales will result in offsets
in Fig.~\ref{fig:YM500}. For example, a number of recent papers have argued that
failure of some of the assumptions underlying the standard methods for
estimating cluster masses from X-ray data (e.g., detailed hydrostatic
equilibrium or the unimportance of turbulent and nonthermal pressure) could
produce a systematic bias in the X-ray cluster mass scale
\citep{planck2011-5.2c,rozo2012b,sehgal2012}.  Finally, as for the LBG sample,
each luminosity bin of the MCXC contains a distribution of halo properties that
are averaged in complicated fashion by our stacked SZ measurement.
Understanding the relative importance of these various effects at a precision
better than 20\% would again require detailed modeling of the heterogeneous MCXC
catalogue.

\section{Conclusions}
\label{sec:conclusions}

Using \Planck\ data, we have measured the scaling relation between
Sunyaev-Zeldovich signal and stellar mass for locally brightest galaxies
($Y_{500}$-$M_\ast$).  This is the first time such a relation has been
determined, and it demonstrates the presence of hot, diffuse gas in halos
hosting central galaxies of stellar mass as low as $M_\ast = 2\times
10^{11}\,\Msolar$, with a strong indication of signal at even lower masses.  We
have constructed a large mock catalogue of locally brightest galaxies from the
Millennium Simulation and used it to model the \Planck\ observational process in
detail in order to extract from our measurements the underlying SZ signal-halo
mass relation ($Y_{500}$-$M_{500}$).  This new relation spans a large range in
halo mass, reaching from rich clusters down to $M_{500}=2.0\times 10^{13}\,
\Msolar$, with a clear indication of continuation to $M_{500}\sim 4\times
10^{12}\,\Msolar$.  This is the lowest mass scale to which an SZ scaling
relation has so far been measured. The fact that the signal is close to the
self-similar prediction implies that \Planck -detected hot gas represents
roughly the mean cosmic fraction of the mass even in such low-mass
systems. Consistency with their low observed X-ray luminosities then requires
the gas to be less concentrated than in more massive systems. Integration of the
halo mass function down to $M_{500}=4\times10^{12}\Msolar$ shows that
\Planck\ has now seen about a quarter of all cosmic baryons in the form of hot
gas, about four times as many as are inferred from X-ray data in clusters with
$M_{500}>10^{14}\Msolar$.

At the high mass end, the scaling relation we derive from our LBG data shows
reasonable agreement with X-ray cluster results.  The 20\% lower normalisation
that we find (significant at the $2.6\sigma$ level) can be explained in
principle by a number of possible effects related to the differing selection and
mass estimation methods of the two samples.  Agreement at this level of
precision is remarkable, and understanding the remaining difference would
require detailed modeling of the selection and calibration of the X-ray samples.
The fact that plausible Malmquist corrections can eliminate most of the
difference shows that cluster studies are now reaching the $\sim10\%$ precision
level.

We find that the $Y_{500}$-$M_{500}$ scaling law is described by a power law
with no evidence of deviation over more than two orders of magnitude in halo
mass.  The gas properties of dark matter halos appear remarkably regular over a
mass range where cooling and feedback processes are expected to vary strongly.
In particular, we find no change in behaviour in the low-mass systems for which
substantial feedback effects are invoked in current galaxy formation models
(e.g., from AGN). Statistical studies of large galaxy and cluster samples, such
as those presented here, can clearly shed new light on the thermal cycle at the
heart of the galaxy formation process.

\begin{acknowledgements}

The authors from the consortia funded principally by CNES, CNRS, ASI, NASA, and
Danish Natural Research Council acknowledge the use of the pipeline-running
infrastructures Magique3 at Institut d'Astrophysique de Paris (France), CPAC at
Cambridge (UK), and USPDC at IPAC (USA).  The development of \Planck\ has been
supported by: ESA; CNES and CNRS/INSU-IN2P3-INP (France); ASI, CNR, and INAF
(Italy); NASA and DoE (USA); STFC and UKSA (UK); CSIC, MICINN, JA and RES
(Spain); Tekes, AoF and CSC (Finland); DLR and MPG (Germany); CSA (Canada); DTU
Space (Denmark); SER/SSO (Switzerland); RCN (Norway); SFI (Ireland); FCT/MCTES
(Portugal); and PRACE (EU). A description of the Planck Collaboration and a list
of its members, including the technical or scientific activities in which they
have been involved, can be found at
\url{http://www.sciops.esa.int/index.php?project=planck}. We acknowledge the use
of the HEALPix package \citep{gorski2005}.

\end{acknowledgements}
\bibliographystyle{aa}
\bibliography{Planck_bib,SZGalaxies}
%
%
\appendix

\section{Robustness of our results to variations in isolation criteria}
\label{ap:1vs2}

As explained in Sect.~\ref{sec:sample}, our locally brightest galaxy catalogue
was built starting from a parent population with $r<17.7$ taken from the
spectroscopic NYU-VAGC and eliminating any candidate with a companion of equal
or brighter $r$~magnitude violating certain isolation criteria. In particular,
we defined locally brightest galaxies to be the set of all objects with $z>0.03$
that are brighter than all other sample galaxies projected within a radius of
$R_{\rm iso} = 1.0$\,Mpc, and differing in redshift by less than
1,000\,km\,s$^{-1}$.  Hereafter, we refer to these criteria as the ``1\,Mpc
case''.

To test the robustness of our results against changes in these isolation
criteria, we compared them to a case with stricter isolation criteria, $R_{\rm
  iso} = 2.0$\,Mpc and 2,000\,km\,s$^{-1}$ in redshift, hereafter the ``2\,Mpc
case''.

Applying the isolation criteria to the parent spectroscopic catalogue as before,
but with these new values, we end up with a first sample of 206,562 locally
brightest galaxes.  Again, in a second step we use SDSS photometry to further
eliminate objects with companions that might violate the isolation
criteria. After removing any candidate with a (photometric) companion of equal
or brighter $r$~magnitude and projected within 2.0\,Mpc, we end up with a
cleaned sample of 110,437 locally brightest galaxies. In particular, this sample
contains 58,105 galaxies satisfying the bound $\log_{10} M_\ast/\rm{\Msolar}\geq
11.0$, which is the regime where we find significant SZ signal. Thus, 23,287
galaxies in this mass range are eliminated from the sample studied in the main
body of this paper by the stricter isolation criteria.

To evaluate the reliability of the new $R_{\rm iso} = 2.0$\,Mpc sample, we
follow the same procedure as before and construct a mock sample of locally
brightest galaxies from the \citet{2011MNRAS.413..101G} simulation. As expected,
the new $R_{\rm iso} = 2.0$\,Mpc mock sample has a higher reliability than the
$R_{\rm iso} = 1.0$\,Mpc case. The fraction of locally brightest galaxies that
are centrals now has a minimum of just over 87\,\% at stellar masses somewhat
above $10^{11}\,\Msolar$. The improvement is less than might have been
anticipated because, as noted in Sect.~\ref{sec:cgc}, the majority of the
satellite galaxies in our simulated 1\,Mpc sample were included because they are
brighter than the central galaxies of their own halos, rather than because the
isolation criteria failed to eliminate them.

\begin{figure*}
\centering
\includegraphics[trim = 2mm 4mm 12mm 8mm, clip=true, width=9cm]{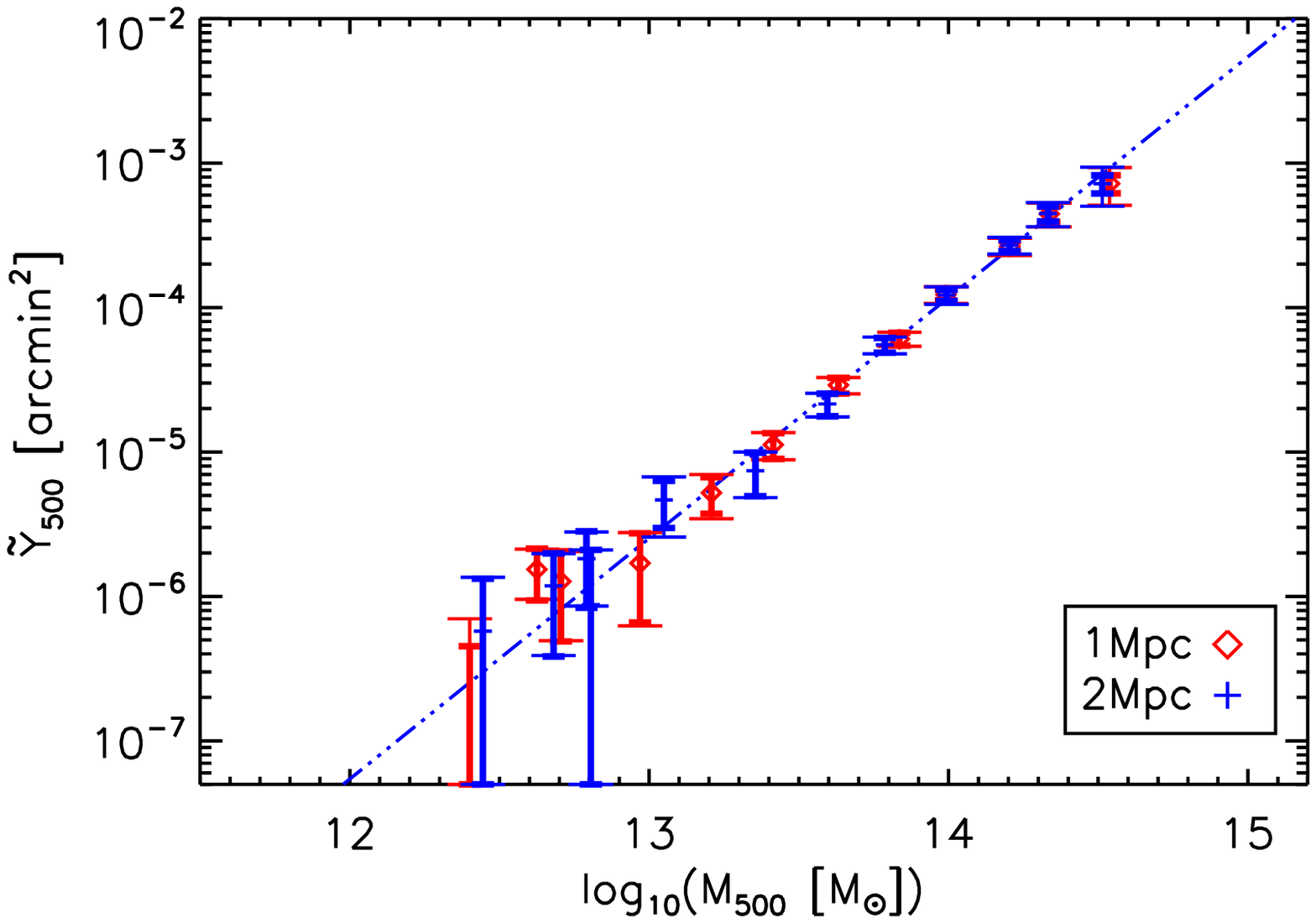}
\includegraphics[trim = 2mm 4mm 12mm 8mm, clip=true, width=9cm]{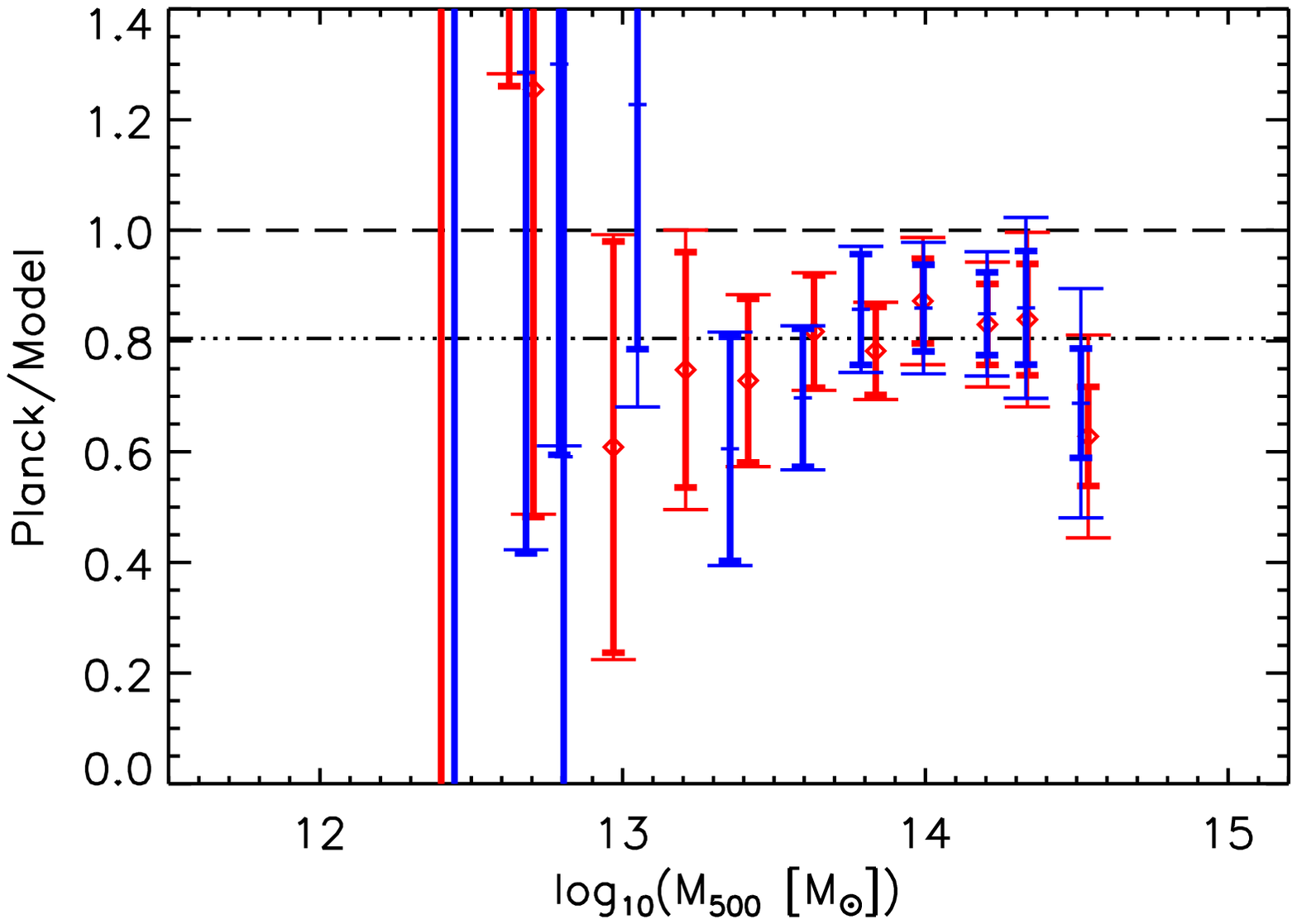}
\caption{ Left: Comparison of the SZ signal-halo mass scaling relation for two
  different sets of isolation criteria.  The triple-dot dashed line is our best
  fit model (see Eq.~\ref{eq:1} and \ref{eq:2}).  Right: Same as above, but now
  showing the ratio of the previous data points to the \citet{arnaud2010}
  $Y$-$M_{500}$ relation. }
\label{fig:1vs2}
\end{figure*}

Finally, Fig.~\ref{fig:1vs2} compares the SZ signal-halo mass relations
($Y_{500}$-$M_{500}$) derived for the two cases (1\,Mpc and 2\,Mpc).  Halo
masses for the 2\,Mpc case are computed as explained in
Sect.~\ref{sec:discussion} (see Table~\ref{tab:bins} for the numerical
values). The main conclusion is that the SZ signal-halo mass scaling relation is
not sensitive to the isolation criteria.

\section{Predicted properties of the stellar mass-halo mass relation}
\label{ap:shm}

\begin{figure*}
\centering
\includegraphics[width=18cm]{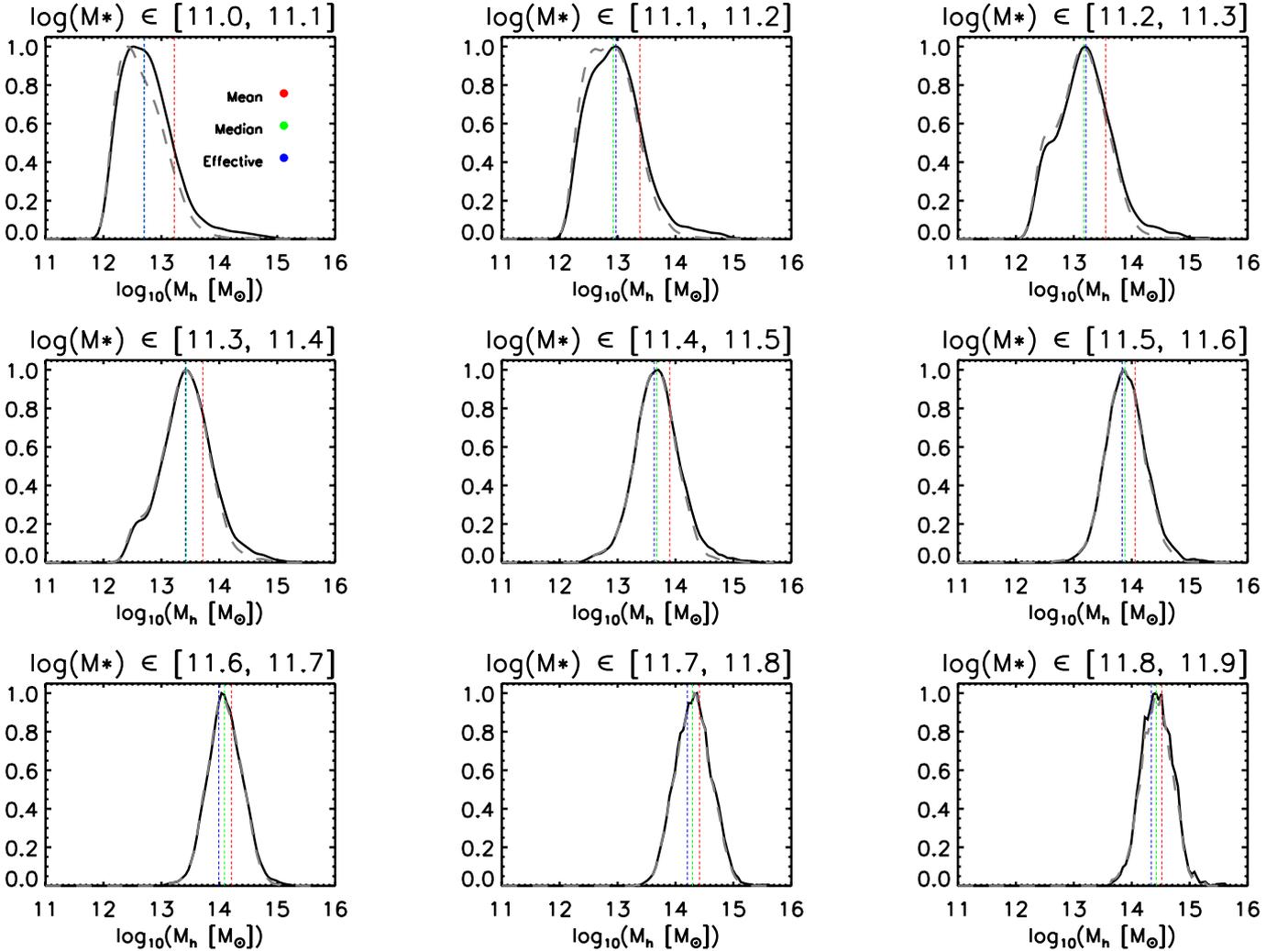}
\caption{ Probability distribution function of halo mass, $\Mhalo$, for nine of
  the stellar mass bins considered in this paper. Solid lines correspond to the
  sample isolated according to the 1\,Mpc criteria, while dashed lines show the
  distributions for the 2\,Mpc sample. Vertical colored lines show three
  different characteristic masses (the mean, median, and ``effective'' halo
  masses) for the 1\,Mpc sample (see Table~\ref{tab:bins} for numerical
  values). }
\label{fig:mhms}
\end{figure*}

Using our mock catalogues based on the semi-analytic galaxy formation simulation
of \citet{2011MNRAS.413..101G}, we provide here additional information on the
predicted properties of the stellar mass-halo mass
relation. Figure~\ref{fig:mhms} shows the distribution of halo mass ($\Mhalo$)
predicted for nine of the stellar mass bins considered in this paper, and for
two sets of isolation criteria: the 1\,Mpc and 2\,Mpc cases (see
Sect.~\ref{sec:sample} and Appendix~\ref{ap:1vs2}). Vertical lines correspond to
the mean (red), median (green), and the ``effective'' (blue) values of halo mass
in each bin. The corresponding numbers are listed in Table~\ref{tab:bins}, which
also gives the RMS of the posterior $\Mhalo$ distribution. The effective halo
masses are computed as described in Sect.~\ref{sec:discussion}.

\begin{table*}[tmb] 
\begingroup 
\newdimen\tblskip \tblskip=5pt
\caption{Statistics of halo mass for various stellar mass bins, for the
  1\,Mpc and 2\,Mpc isolation criteria. The first three columns for each
  case (mean, median, and RMS values for the halo mass) are derived from the
  simulation only, while the effective halo mass $\Mhalo^{\rm eff}$ uses the
  redshifts and stellar masses of the observed galaxies, as described in
  Sect.~\ref{sec:discussion}. All masses ($M$) in this table are decimal
  logarithms of the value in units of $\Msolar$.
}
\label{tab:bins}
\vskip -3mm
\footnotesize 
\setbox\tablebox=\vbox{ %
\newdimen\digitwidth 
\setbox0=\hbox{\rm 0}
\digitwidth=\wd0
\catcode`*=\active
\def*{\kern\digitwidth}
\newdimen\signwidth
\setbox0=\hbox{+}
\signwidth=\wd0
\catcode`!=\active
\def!{\kern\signwidth}
\halign{\hbox to 2.5cm{#\leaderfil}\tabskip=2em&
     \hfil#\hfil\tabskip=0.4em&
     \hfil#\hfil&
     \hfil#\hfil&
     \hfil#\hfil\tabskip=2em&
     \hfil#\hfil\tabskip=0.4em& 
     \hfil#\hfil&
     \hfil#\hfil&
     \hfil#\hfil&
     \hfil#\hfil\tabskip=0pt\cr
\noalign{\doubleline}
\omit&\multispan8\hfil$\log_{10}\left({M_{\rm h}\over \Msolar}\right)$\hfil\cr
\noalign{\vskip -3pt}
\omit&\multispan8\hrulefill\cr
\noalign{\vskip 2pt}
\omit&\multispan4 \hfil $R_{\rm iso} = 1$\,Mpc \hfil & \multispan4\hfil $R_{\rm iso} = 2$\,Mpc \hfil\cr
\noalign{\vskip -3pt}
\omit& \multispan4\hrulefill&\multispan4\hrulefill\cr
\noalign{\vskip 3pt}
\omit\hfil$\log_{10}\left({M_\ast\over\Msolar}\right)$\hfil&
     \omit\hss Mean\hss&\omit\hss Median\hss&\omit\hss RMS\hss&\omit\hss Effective\hss&
      Mean&Median& RMS&\omit\hss Effective\hss\cr
\noalign{\vskip 3pt\hrule\vskip 5pt}
11.0--11.1& 13.22& 12.70& 13.86& 12.71& 12.92& 12.61& 13.39& 12.79\cr
11.1--11.2& 13.38& 12.93& 13.94& 12.97& 13.14& 12.85& 13.67& 12.81\cr
11.2--11.3& 13.55& 13.17& 14.04& 13.21& 13.37& 13.12& 13.79& 13.05\cr
11.3--11.4& 13.72& 13.43& 14.03& 13.41& 13.60& 13.40& 13.87& 13.35\cr
11.4--11.5& 13.90& 13.67& 14.15& 13.63& 13.81& 13.65& 13.92& 13.60\cr
11.5--11.6& 14.06& 13.89& 14.19& 13.84& 14.01& 13.87& 14.10& 13.79\cr
11.6--11.7& 14.21& 14.09& 14.19& 13.99& 14.19& 14.08& 14.13& 13.99\cr
11.7--11.8& 14.41& 14.29& 14.39& 14.20& 14.39& 14.29& 14.25& 14.20\cr
11.8--11.9& 14.52& 14.42& 14.49& 14.34& 14.49& 14.42& 14.30& 14.33\cr
11.9--12.0& 14.71& 14.60& 14.56& 14.54& 14.69& 14.60& 14.52& 14.51\cr
\noalign{\vskip 5pt\hrule\vskip 3pt}}}
\endPlancktable 
\endgroup
\end{table*}

\section{Impact of miscentering and scatter on the binned SZ signal and
stacked SZ maps}
\label{ap:miscenter}

\begin{figure}
\centering
\includegraphics[width=9cm]{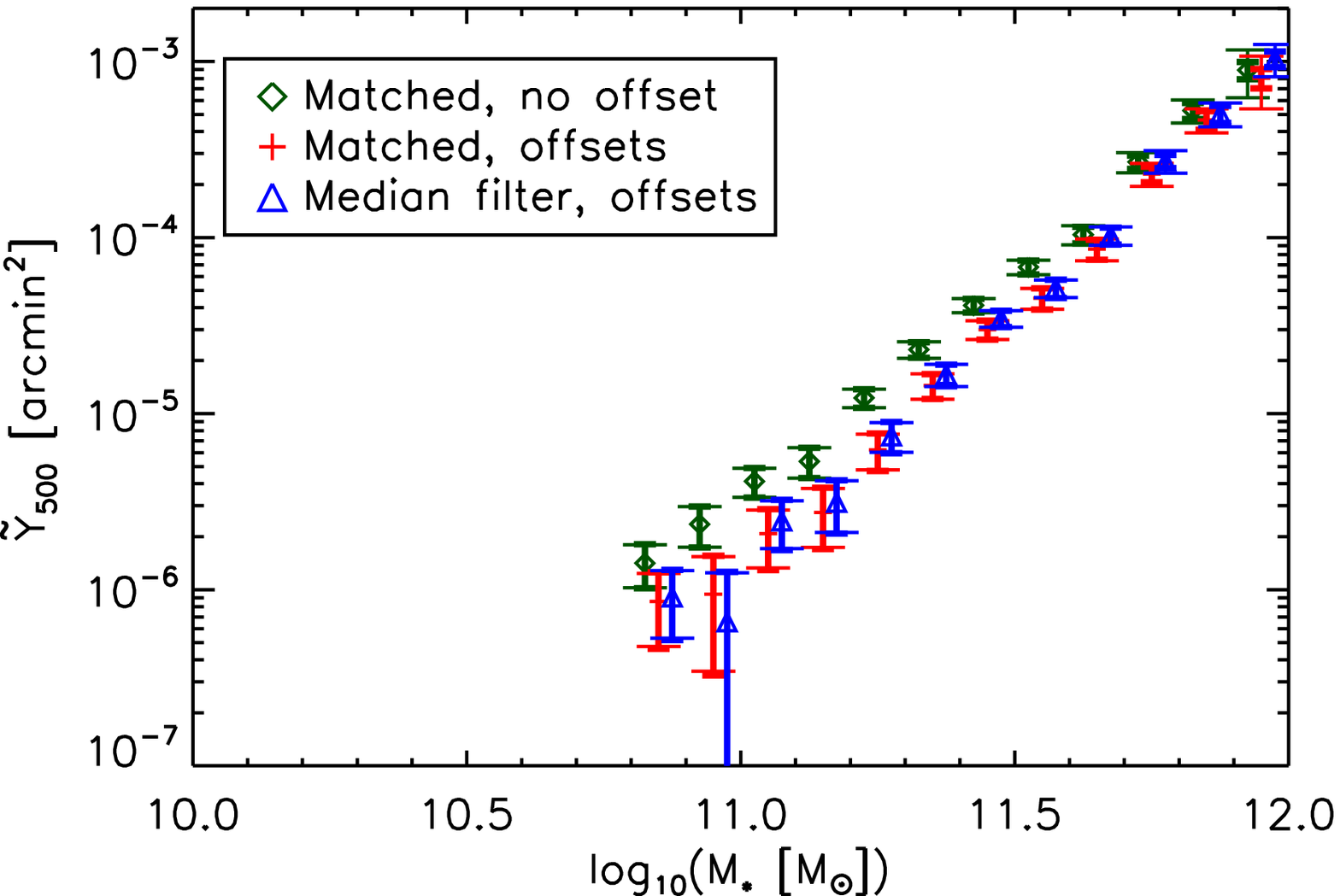}
\caption{Impact of miscentering and scatter on the binned SZ measurements.  The
  green points give results in an ideal situation with no miscentering and SZ
  filter perfectly matched to each individual object in a given stellar mass
  bin. The red crosses add the effect of miscentering, with offsets drawn from
  the distributions given by the simulations for each stellar mass bin. The blue
  triangles additionally include the aperture effect caused by fixing the filter
  size according to the median value of the halo mass in each bin.}
\label{fig:Yimpact}
\end{figure}

As discussed in Sec.~\ref{sec:discussion}, we used the semi-analytic galaxy
formation simulation of \citet{2011MNRAS.413..101G} to account for the effects
of miscentering and scatter in halo mass at fixed stellar mass when interpreting
our measurement (see Figs.~\ref{fig:Ybinned} and \ref{fig:YM500}).
Figure~\ref{fig:Yimpact} isolates the impact of each effect on the binned SZ
measurements, using the procedure outlined in that section.  The green points
represent the ideal case with no miscentering and SZ filter perfectly matched to
the size of each individual object.  The red crosses add miscentering offsets
taken from the offset distribution in the simulations for each stellar mass bin.
The drop in SZ amplitude is expected because we now miss SZ signal from the
miscentered objects.  Additionally fixing the filter size according to the
median halo mass in each stellar mass bin, as done throughout this paper, we
recover our previous results, shown as the blue triangles here and as the red diamonds
in the left-hand panel of Fig.~\ref{fig:YM500}.

\begin{figure*}
\centering
\includegraphics[width=14.5cm]{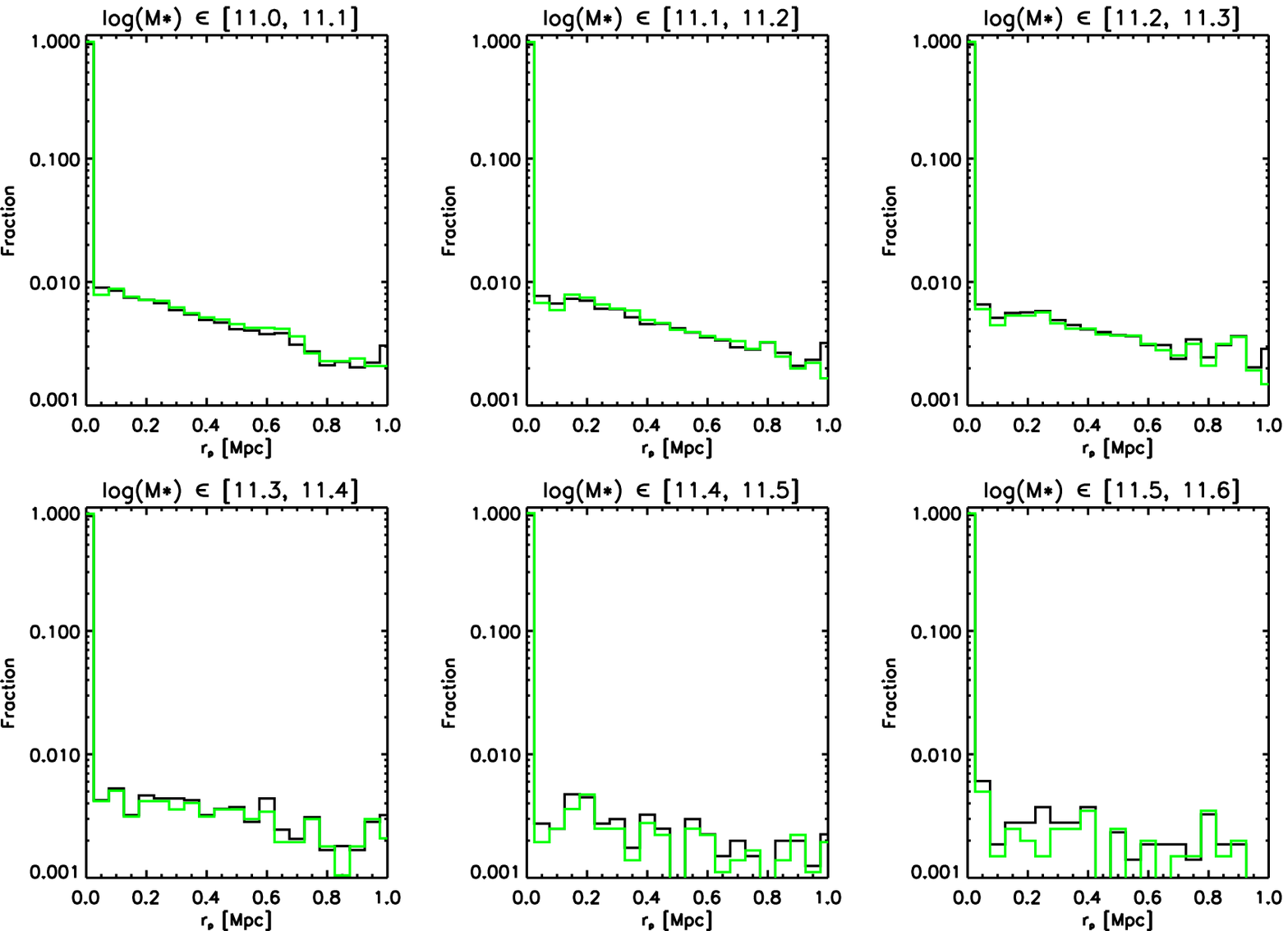}
\caption{ Distribution of offsets of locally brightest galaxies from the
  gravitational potential minima of their parent halos, both for the 1\,Mpc
  (black) and for the 2\,Mpc (green) isolation criteria. Table~\ref{tab:rpbins}
  gives mean and RMS values for these distributions. }
\label{fig:historp}
\end{figure*}

Using the same simulations, we can also estimate the impact of miscentering on
the stacked SZ maps of locally brightest galaxies (see
Fig.~\ref{fig:stacking-y-map}).  Here, we use the full simulation to compute
$r_{\rm p}$, the projected distance of each locally brightest galaxy from the
gravitational potential minimum of its halo.  Average and RMS values for $r_{\rm
  p}$ for all the stellar mass bins considered in this paper and for the 1\,Mpc
and 2\,Mpc samples are given in Table~\ref{tab:rpbins}. Histograms of these
$r_{\rm p}$ values are shown in Fig.~\ref{fig:historp}. Note that the median
value of $r_{\rm p}$, which is not listed in the table, is zero for all bins.

\begin{table}[tmb] 
\begingroup 
\newdimen\tblskip \tblskip=5pt
\caption{Statistics of the distribution of distances $r_{\rm p}$ of the locally
  brightest galaxies from the gravitational potential minima of their
  parent halos, for the 1\,Mpc and 2\,Mpc isolation criteria. 
}
\label{tab:rpbins}
\nointerlineskip
\vskip -3mm
\footnotesize 
\setbox\tablebox=\vbox{ %
\newdimen\digitwidth 
\setbox0=\hbox{\rm 0}
\digitwidth=\wd0
\catcode`*=\active
\def*{\kern\digitwidth}
\newdimen\signwidth
\setbox0=\hbox{+}
\signwidth=\wd0
\catcode`!=\active
\def!{\kern\signwidth}
\halign{\hbox to 2.5cm{#\leaderfil}\tabskip=2em&
    \hfil#\hfil\tabskip=1em&
    \hfil#\hfil\tabskip=2em& 
    \hfil#\hfil\tabskip=1em&
    \hfil#\hfil\tabskip=0pt\cr
\noalign{\doubleline}
\omit&\multispan4\hfil$r_{\rm p}$\,[kpc]\hfil\cr
\noalign{\vskip -3pt}
\omit&\multispan4\hrulefill\cr
\noalign{\vskip 4pt}
\omit&\multispan2\hfil R$_{\rm iso}=1$\,Mpc\hfil&\multispan2\hfil R$_{\rm iso}=2$\,Mpc\hfil\cr
\noalign{\vskip -3pt}
\omit& \multispan2\hrulefill& \multispan2\hrulefill\cr
\noalign{\vskip 1pt}
\omit\hfil$\log_{10}\left({M_\ast\over\Msolar}\right)$\hfil&Mean&RMS&Mean&RMS\cr
\noalign{\vskip 3pt\hrule\vskip 5pt}
11.0--11.1& 140.2& *469.6&  *75.8& *322.1\cr
11.1--11.2& 165.7& *533.1&  *94.9& *373.3\cr
11.2--11.3& 195.6& *636.1&  121.8& *501.8\cr
11.3--11.4& 202.4& *682.1&  143.6& *579.1\cr
11.4--11.5& 217.8& *720.3&  165.1& *659.8\cr
11.5--11.6& 239.8& *852.8&  205.7& *812.1\cr
11.6--11.7& 193.4& *775.2&  171.5& *758.5\cr
11.7--11.8& 213.4& *896.7&  200.4& *892.0\cr
11.8--11.9& 145.1& *726.2&  128.5& *720.9\cr
11.9--12.0& 342.5& 1062.6&  332.2& 1065.1\cr
\noalign{\vskip 5pt\hrule\vskip 3pt}}}
\endPlancktable 
\endgroup
\end{table}

\begin{figure*}
\centering
\includegraphics[width=14.5cm]{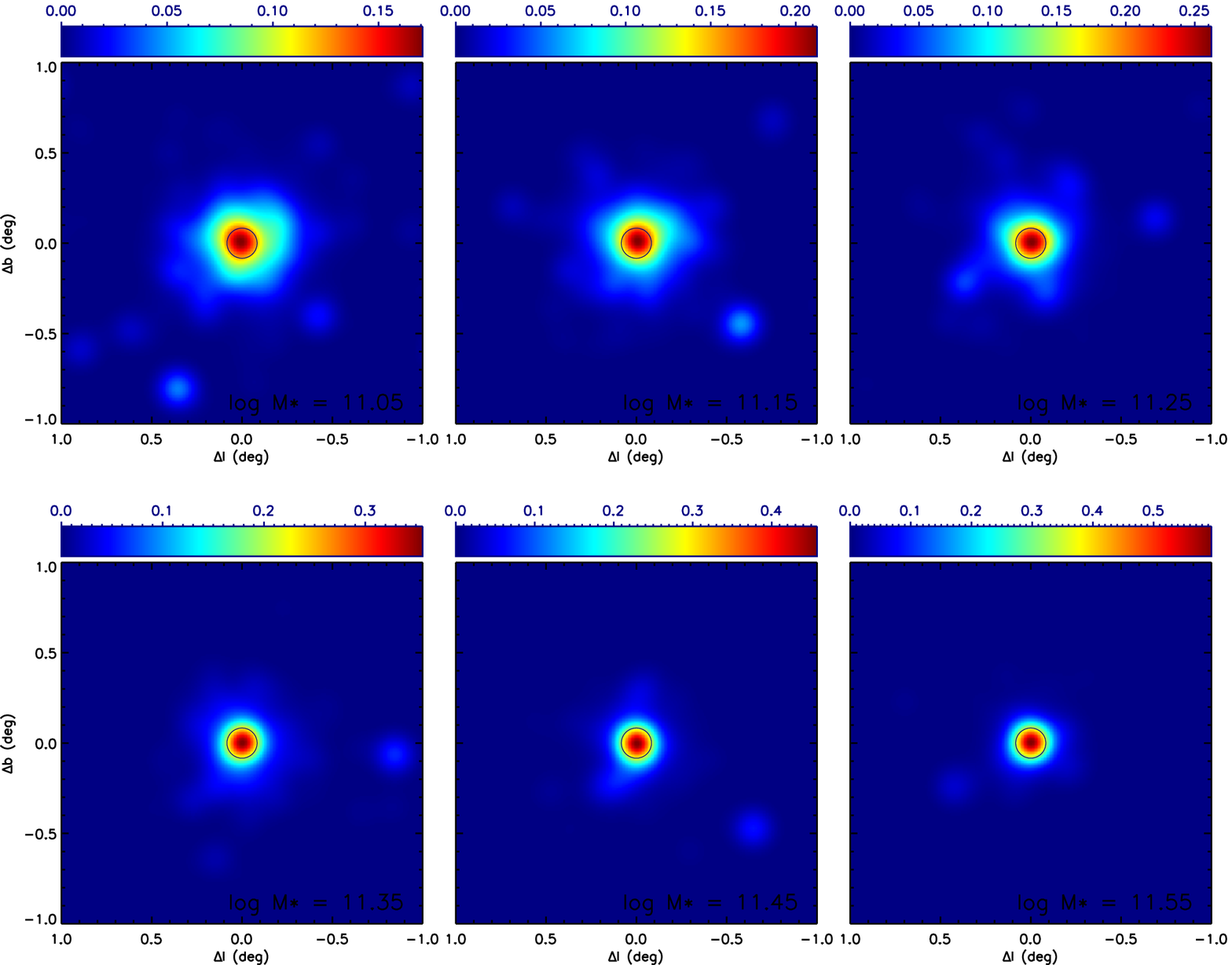}
\caption{ Impact of miscentering on stacked SZ maps. See the text for details of
  the simulation shown here. For an original resolution of FWHM$=10\arcm$,
  miscentering broadens the stacked profiles to a FWHM$\sim 20\arcm$ for
  $\log_{10} M_\ast/\rm{\Msolar} \leq 11.25$.}
\label{fig:miscenter}
\end{figure*}

These values can be used to predict the impact of miscentering of the locally
brightest galaxy with respect to its halo (and thus, with respect to the centre
of the associated SZ emission). Figure~\ref{fig:miscenter} illustrates the
broadening of the SZ stacked profile caused by this effect. For this
computation, we assume point-like objects and a Gaussian beam profile of
$10\arcm$ for easier comparison with Fig.~\ref{fig:stacking-y-map}. For each
stellar mass bin, the $\Mhalo$ value from the simulation is used to predict the
total SZ flux using Eqs.~\ref{eq:1} and \ref{eq:2}, and the $r_{\rm p}$ value is
used to offset the position of the SZ signal. In order to convert $r_{\rm p}$
values (in physical units) into angular offsets, a redshift for each simulated
object is drawn from the observed distribution for locally brightest galaxies of
similar stellar mass.  Miscentering broadens the stacked SZ profile, yielding
typical FWHM of $\sim 20\arcm$ for $\log_{10} M_\ast/\rm{\Msolar} \leq 11.25$,
and also modifies the shape of the profile, by increasing the amount of SZ flux
in the tails of the distribution. These values are slightly smaller (but
comparable) to the observed widths of the SZ emission in
Fig.~\ref{fig:stacking-y-map}.

\begin{figure*}
\centering
\includegraphics[width=14.5cm]{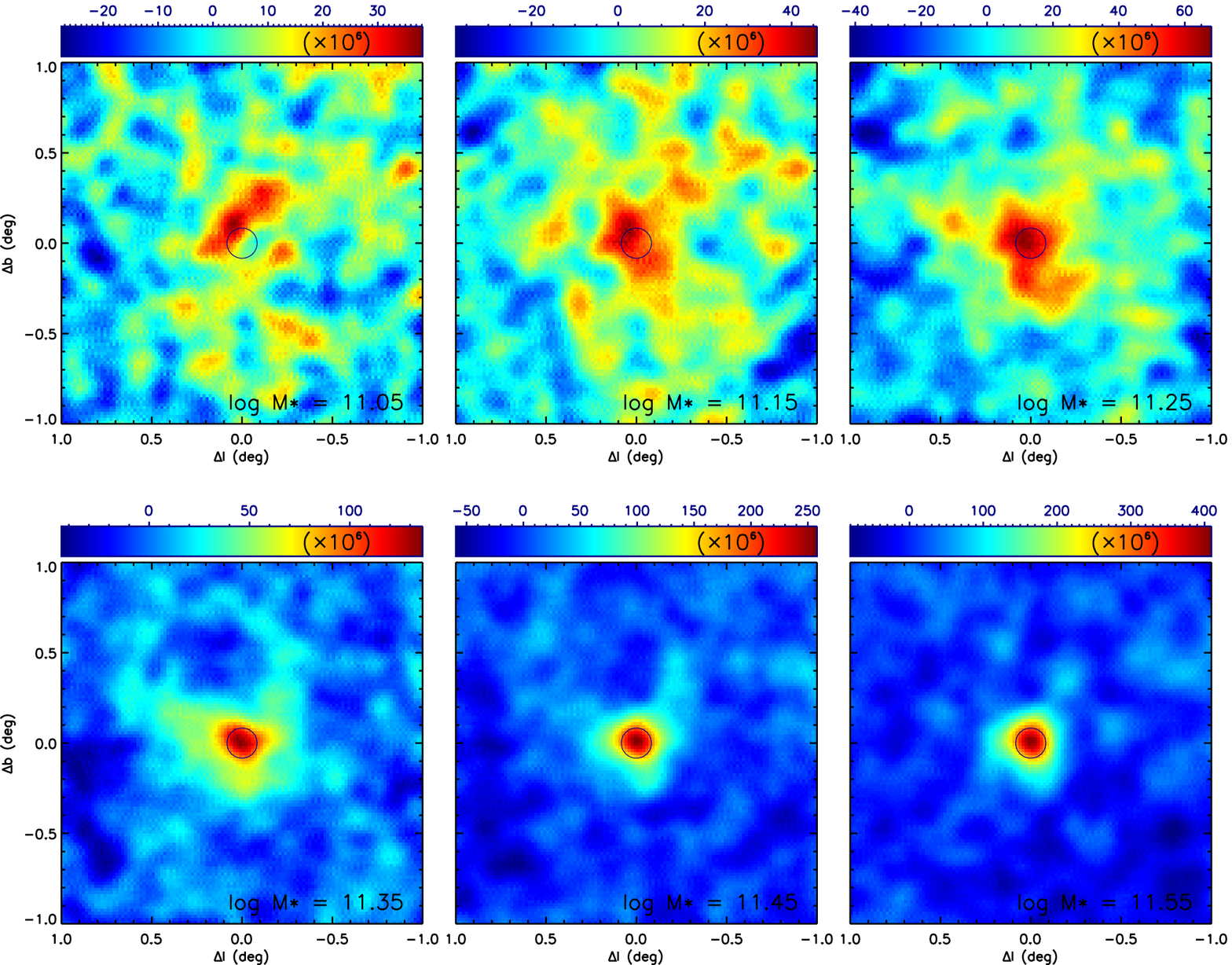}
\caption{Similar to Fig.~\ref{fig:stacking-y-map}, but using a reconstructed SZ
  map that now uses all six HFI frequency channels. The noise in all maps is
  reduced by the inclusion of the two highest frequencies. Stacked images in the
  stellar-mass bins above $\log_{10}(M_\ast/\Msolar) = 11.25$ are not
  significantly affected, but for the low stellar-mass panels, the extended
  signal near map centre is larger and has higher signal to noise than in
  Fig.~\ref{fig:stacking-y-map}, suggesting that it may be real SZ signal
  broadened by miscentering effects.}
\label{fig:stacking-y-map2}
\end{figure*}

Finally, Fig.~\ref{fig:stacking-y-map2} shows equal-weighted stacks of SZ maps
centred on the real central galaxy sample, similar to those of
Fig.~\ref{fig:stacking-y-map}, but now using all six HFI frequency channels in
the MILCA algorithm, rather than just the lowest four. For all six stellar mass
bins the noise in these new maps, as measured by the RMS fluctuation about the
mean in pixels more than 20\arcm\ from map centre, is lower than in the maps of
Fig.~\ref{fig:stacking-y-map}.  This shows that the addition of high frequency
information has improved the accuracy with which non-SZ signals, primarily dust
emission, are removed.  Almost all this improvement comes from the inclusion of
the 545\,GHz channel; maps made with and without the 857\,GHz channel are almost
identical. As a result of this improvement, the signal-to-noise ratio of the
peaks near the map centre is higher in all the panels of
Fig.~\ref{fig:stacking-y-map2} than in the corresponding panels of
Fig.~\ref{fig:stacking-y-map}. This strengthens our conclusion that the apparent
SZ signals near the centres of the two lowest stellar mass panels are, in fact,
real, despite their apparent breadth and irregularity. The breadth is likely due
to the miscentering effects explored above while the irregularity looks
consistent with the overall noise level of the maps.

\raggedright
\end{document}